\documentclass[a4paper, 12pt]{article}

\usepackage[
 margin=1.0in,
 top = .45in,
  bottom = .44in,
  includefoot,
  footskip=30pt,
]{geometry}

\usepackage{todonotes}

\usepackage{setspace}
\usepackage[pagewise]{lineno}

\usepackage{amsfonts,amsmath,amssymb,amsthm}
\usepackage{graphicx,subcaption, epsfig,psfrag}
\usepackage{array, booktabs, placeins}
\usepackage{bm,dsfont,color,appendix}
\usepackage{natbib}
\usepackage{latexsym,graphicx}
\usepackage{hyperref}
\usepackage{bbold}

\usepackage{algorithm}
\usepackage{algpseudocode}
\newtheorem{theorem}{Theorem}
\newtheorem{lemma}{Lemma}
\newtheorem{remark}{Remark}

\theoremstyle{plain}

\newcommand{\argmin}{\mathop{\mathrm{argmin}}}
\newcommand{\betapre}{{\hat \beta_{\mathrm{pre}}}}
\newcommand{\signb}{\mathbf{sign}}
\def\half{\frac{1}{2}}

\begin{document}
\title{Pretraining and the Lasso}

\author{Erin Craig$^{1}$  \and  Mert Pilanci$^2$ \and Thomas Le Menestrel$^3$ \and Balasubramanian Narasimhan$^4$ \and Manuel A. Rivas$^1$\and Stein-Erik Gullaksen$^{6, 7}$ \and Roozbeh Dehghannasiri$^1$ \and Julia Salzman$^{1,5}$ \and Jonathan Taylor$^4$  \and
 Robert Tibshirani$^{1,4}$}
\date{\footnotesize
$^1$Department of Biomedical Data Science, Stanford University, Stanford CA, USA \\
$^2$Department of Electrical Engineering, Stanford University, Stanford CA, USA \\
$^3$ Institute for Computational and Mathematical Engineering, Stanford University, Stanford CA, USA \\
$^4$Department of Statistics, Stanford University, Stanford CA, USA \\
$^5$ Department of Biochemistry, Stanford University, Stanford CA, USA \\
$^6$ Department of Medicine, Hematology Section, Haukeland University Hospital, Bergen, Norway\\
$^7$ K.G. Jebsen Centre for Myeloid Blood Cancer, Department of Clinical Science, University of Bergen, Bergen, Norway
}



\maketitle
\begin{abstract}
    Pretraining is a popular and powerful paradigm in machine learning to pass information from one model to another. As an example, suppose one has a modest-sized dataset of images of cats and dogs, and plans to fit a deep neural network to classify them from the pixel features. With pretraining, we start with a neural network trained on a large corpus of images, consisting of not just cats and dogs but hundreds of other image types. Then we fix all of the  network weights except for the top layer(s) (which makes the final classification) and train (or ``fine tune'') those weights on our dataset. This often results in dramatically better performance than the network trained solely on our smaller dataset.
In this paper, we ask the question ``Can pretraining help the lasso?''.  We develop a framework for the lasso in which a model is fit to a large dataset, and then fine-tuned using a smaller dataset. This latter dataset can be a subset of the original dataset, or it can be a dataset with a different but related outcome. This framework has a wide variety of applications, including stratified models, multinomial responses, multi-response models, conditional average treatment estimation and even gradient boosting. 
In the stratified model setting, the pretrained lasso pipeline estimates the coefficients common to all groups at the first stage, and then group-specific coefficients at the second ``fine-tuning'' stage. We show that under appropriate assumptions, the support recovery rate of the common coefficients is superior to that of the usual lasso trained only on individual groups. This separate identification of common and individual coefficients can also be useful for scientific understanding.

\end{abstract}


\section{Introduction}
\label{sec:intro}
Pretraining is a popular and powerful tool in machine learning. As an example, suppose you want to build a neural net classifier to discriminate between images of cats and dogs, and suppose you have a labelled training set of say 500 images.
You could train your model on this dataset, but a more effective approach 
is to start with a neural net trained on a much larger corpus of images, for example IMAGENET (1000 object classes and 1,281,167 training images).
The weights in this fitted network are then fixed, except for the top layers which
make the final classification of dogs vs cats; finally, the weights in the top layers are refitted using our training set of 500 images. This approach is effective because the network pretrained on a large corpus can discover potentially predictive features for our discrimination problem.  
This paper asks: is there a version of pretraining for the lasso? We propose such a framework.

This work was motivated by a study carried out with Genentech \citep{mcGough2023}.
The authors curated a pancancer dataset, consisting of  10 groups of patients with different cancers. Some of the cancer classes are large  (e.g. breast, lung) and some are smaller (e.g. head and neck).   
The goal is to predict survival times from a large number of features (labs, genetics, $\ldots$), approximately 2,000 in total. They compare two approaches: (a) a ``pancancer model'', in which a single model is fit to the training set and used  to make predictions for all cancer classes and (b) separate (class specific) models, each used to make predictions for one class. The authors found that the two approaches produced very similar results, with the pancancer model offering a small advantage in test set C-index for the smaller classes (e.g. head and neck cancer). Presumably this occurs because of the insufficient sample size for fitting a separate head and neck cancer model, so that ``borrowing strength'' across a set of different cancers can be helpful. 

This led us to consider a framework where the pancancer model is adaptively blended with individual models, allowing each to ``learn'' from the pancancer model while identifying effects specific to each group. \emph{Importantly, this framework is not specific to grouped data; rather it can be applied to any setting where we wish to share information from one model to another.} This paradigm is somewhat closely related to the ML pretraining mentioned above. It also has similarities to {\em transfer learning}.

This paper is organized as follows. In Section \ref{sec:pretraining} we review the lasso, describe the pretrained lasso, and show an example with real data. Section \ref{sec:related} discusses
related work. In Section \ref{sec:useCases} we demonstrate the generality of the idea, detailing a number
of different ``use cases'' in Sections \ref{sec:multinomial_response} -- \ref{sec:CATE}. We move beyond linear models in Section \ref{sec:boost}, illustrating an application of the pretrained lasso to gradient boosting. Real data examples are shown throughout the paper, including applications to cancer, genomics, and chemometrics. Section \ref{sec:theory} establishes some theoretical results for the pretrained lasso. In particular we show that under the 
``shared/ individual model'' discussed earlier, the new procedure enjoys improved rates of support recovery, as compared to the usual lasso. We end with a discussion in Section \ref{sec:discussion}. 

\section{Pretraining the lasso} 
\label{sec:pretraining}

\subsection{Review of the lasso}
For the Gaussian family with data $(x_i,y_i), i=1,2,\ldots n$, the lasso has the form 
\begin{equation}
{\rm argmin}_{\beta_0, \beta} \half\sum_{i=1}^n(y_i- \beta_0 -\sum_{j=1}^p x_{ij}\beta_j)^2 + \lambda \sum_{j=1}^p |\beta_j |.
\end{equation}
Varying the  regularization parameter $\lambda \ge 0$ yields a path of solutions: an optimal value
$\hat\lambda$ is usually chosen by cross-validation, using for example the {\tt cv.glmnet} function in the R language package {\tt glmnet}~\citep{glmnet}.

 Before presenting our proposal, two more background facts are needed. In GLMs and $\ell_1$-~regularized GLMs, one can include an \emph{offset}:
 this is a pre-specified $n$-vector that is included as an additional column to the feature matrix, but whose weight $\beta_j$ is fixed at 1. Secondly, one can generalize the $\ell_1$ norm $\sum_j |\beta_j |$ to a \emph{weighted} norm
 $\sum_j {\rm pf}_j |\beta_j |$
where each ${\rm pf}_j \ge 0$ is a {\em penalty factor} for feature $j$. At the extremes, a penalty factor of zero implies no penalty and means that the feature will always be included in the model; a penalty factor of $+\infty$ leads to that feature being discarded. 

\subsection{The algorithm}
\label{sec:alg}

Suppose we express our data as
an  $N \times p$  feature matrix $X$ and a target $N$-vector $y$,
and we want to do supervised learning via the lasso. In the training set, suppose further that each observation falls in one of $K$ pre-specified classes, and therefore the rows of our data are partitioned into groups $X_1, \dots, X_{K}$ and $y_1, \dots, y_K$. 

\begin{figure}
    \centering
    \includegraphics[width=4in]{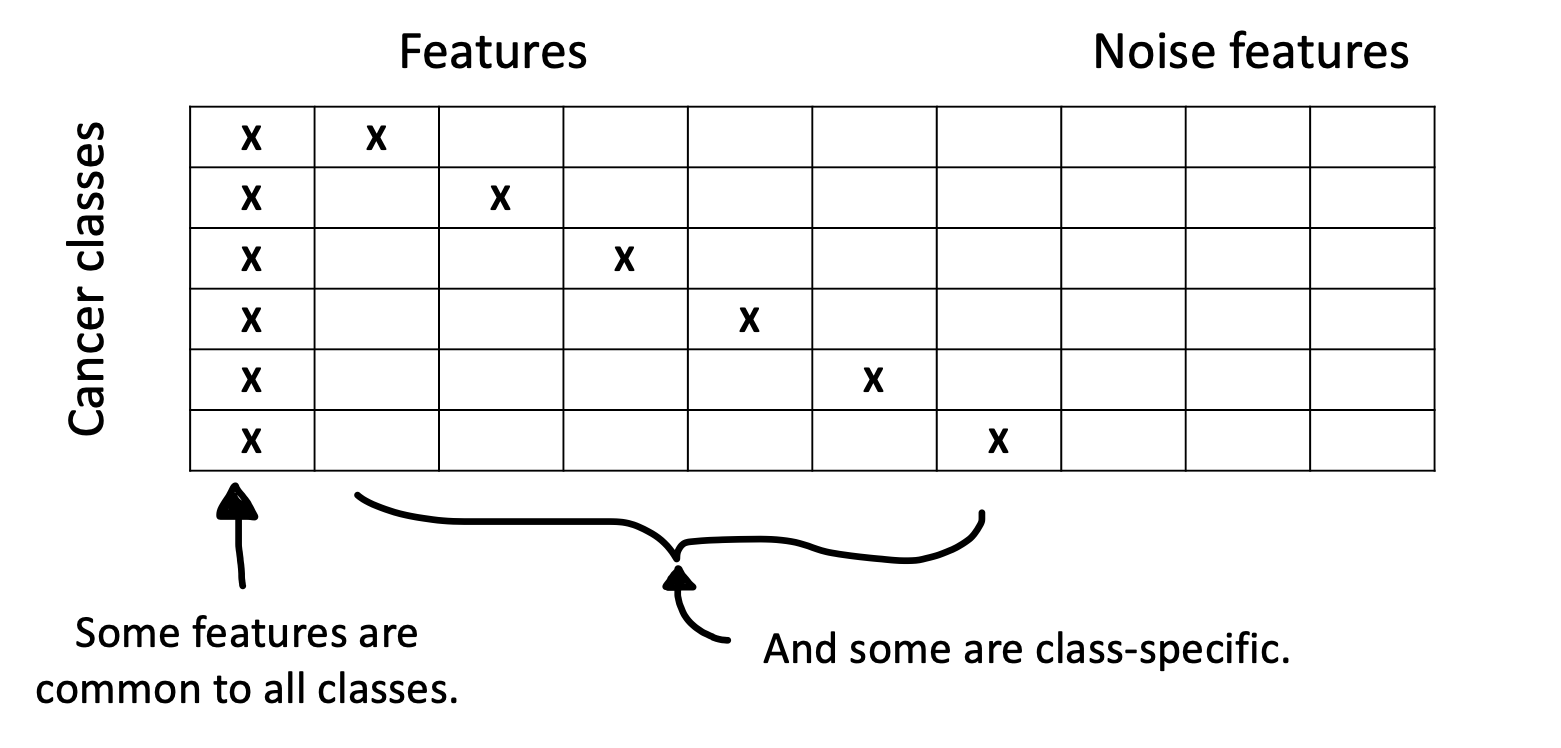}
    \caption{\em Conceptual model: some features are predictive for all or most classes, some are specific to each class, and some are noise.}
    \label{fig:conceptual}
\end{figure}
As shown in Figure \ref{fig:conceptual}, we imagine that the features are roughly divided into three types: \emph{common features} that are predictive in most or all classes, \emph{individual features}, predictive in one particular class and \emph{noise features} with little or no predictive power. Note that the common features may have different effect sizes across classes. Our proposal for this problem is a two-step procedure, with the first step aimed at discovering the common features and the second step focused on recovery of the individual features.

For simplicity, we assume here that $y$ is a Gaussian response ($y$ can also be any member of the GLM family, such as  binomial, multinomial, or Cox survival). Our model has overall  mean and slope components 
$\mu_0$ and $\beta_0$, and class-wise means and slopes:
$\mu_k, \beta_k, k=1,2,\ldots K$.

\begin{equation}
y_k = (\mu_0 + \mu_k) + X_k(\beta_0+ \beta_k) + \varepsilon_k\; \;{\rm  for,}\; k=1,2,\ldots K,
\end{equation}
where $\left(X_k, y_k\right)$ is the subset of observations in group $k$. Note that $\beta_0$ is shared across all classes $k$; this is intended to capture the common features. Then $\beta_k$ captures features that are unique to each class, and may additionally adjust the coefficient values in $\beta_0$. Note that the parameters are not all identifiable as stated: but as we will see, we use an $\ell_1$ penalty in its estimation, which makes them identifiable.

We fit this model in two steps, the first of which is aimed at discovering $\beta_0$, the coefficients shared across groups. We train an \textit{overall} model using all the data:
\begin{equation}
 	\hat{\mu}_0, \hat{\beta}_0 = \argmin_{\mu, \beta} \frac{1}{2} \sum_{k=1}^K \| y_k - \left(\mu \mathbf{1} + X_k \beta\right) \|_2^2 + \lambda ||\beta||_1,
\end{equation}
for some choice of $\lambda$ (e.g the value minimizing the CV error). Define $S(\hat\beta_0)$ to be the support set (the nonzero coefficients) of $\hat{\beta}_0$. 

Now for each group $k$, we  fit a \textit{class specific} model aimed at discovering class specific features $\beta_k$: we find $\hat{\beta}_k$ and $\hat{\mu}_k$ such that
\begin{eqnarray}
&& \hat{\mu}_k, \hat{\beta}_k = \argmin_{\mu, \beta} \frac{1}{2}  \| y_k - (1-\alpha) \left(\hat{\mu}_0 \mathbf{1} + X_k \hat{\beta}_0\right) - (\mu \mathbf{1} + X_k \beta) \|_2^2 +
\cr && \phantom{\hat{\mu}_k, \hat{\beta}_k = } \lambda \sum_{j=1}^p \Bigl[ I(j \in S(\hat{\beta}_0))+ \frac{1}{\alpha} I(j \notin S(\hat{\beta}_0))  \Bigr] |\beta_{j}|.
\label{eq:model}
\end{eqnarray}
We choose $\lambda$ through cross-validation, and $\alpha\in [0,1]$ is a hyperparameter. The class specific models (with the offset from the overall model) are then used for prediction in each group.

When $\alpha=0$, this very nearly returns the overall model, and when $\alpha=1$ this is equivalent to fitting a class specific model for each class. This property is the result of the inclusion of two terms that interact with $\alpha$. 

First, the offset $(1-\alpha) \left(\hat{\mu}_0 \mathbf{1} + X_k \hat{\beta}_0\right)$ in the loss determines how much the prediction from the overall model influences the class specific models. When the response is Gaussian, using this term is the same as fitting a residual: the target is ${y_k - (1-\alpha) \left(\hat{\mu}_0 \mathbf{1} + X_k \hat{\beta}_0\right)}$. That is, the class specific model can only find signal that was left over after taking out the overall model's contribution. When $\alpha = 0$, the class specific model is forced to use the overall model, and when $\alpha = 1$, the overall model is ignored.

Second, the usual lasso penalty is modified by a penalty factor that is $1$ for $\beta_j$ in the support of the overall model $S(\hat{\beta}_0)$ and $\frac{1}{\alpha}$ off the support.
 When $\alpha = 0$, the class specific model is only able to use features in the support of the overall model: the penalty factor is $\infty$ off the support. When $\alpha = 1$, the penalty factor is $1$ everywhere, and all variables are penalized equally as in the usual lasso. 
\begin{remark}
\normalfont
\small
In  our numerical experiments and theoretical analysis (Section \ref{sec:theory}), we find that the transmission of both ingredients--- the offset and penalty factor---
are important for the success of the method.
The offset captures the first stage model, while the penalty factor captures its support. 
\end{remark}

\begin{figure}
    \centering
    \includegraphics[width=.8 \linewidth]{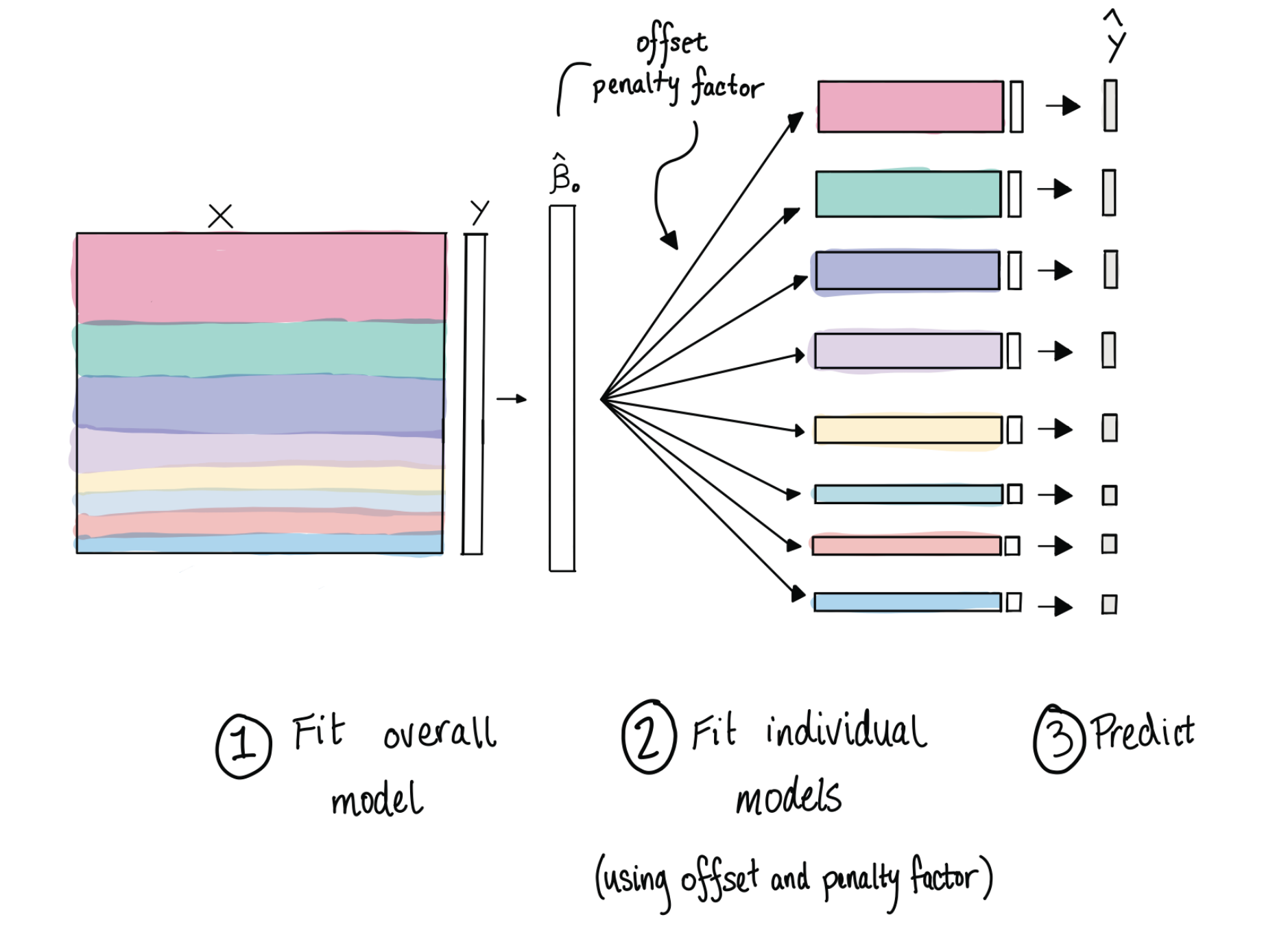}
    \caption{\em Workflow for the pretrained lasso, as applied to grouped data.}
    \label{fig:roadmap}
\end{figure}

The  pretrained lasso algorithm is summarized in Algorithm \ref{alg:alg1}. For clarity we express the computation in terms of the R language package {\tt glmnet}, although in principal this could be any package for fitting $\ell_1$-regularized  generalized linear models and the Cox survival model. A roadmap of the procedure is in Figure \ref{fig:roadmap}. As we will describe in Section \ref{sec:useCases},
our proposed paradigm is far more general: we first describe the simplest case for ease of exposition.



\begin{algorithm}[ht]
\caption{Pretrained lasso with fixed input groups}
  \label{alg:alg1}
\begin{enumerate}
\item Fit a single (``overall'') lasso model to the training set, using for example {\tt cv.glmnet} in the R language. From this, choose a model (weight vector) $\hat\beta_0$ along the $\lambda$ path, using e.g. lambda.min --- the value minimizing the CV error.
\item Fix $\alpha \in [0, 1]$.  Define the {\tt offset} and {\tt penalty factor} as follows:
\begin{itemize}
    \item Define ${{\tt offset}=(1-\alpha)\cdot (X_k\hat\beta_0 + \hat\mu_0)}$.
    \item Let $S$ be the support set of $\hat\beta_0$. Define the penalty factor ${\tt pf}$ as ${{\tt pf}_j=  I(j \in S) + \frac{1}{\alpha} \cdot I(j \notin S)}$.
\end{itemize}
For each class $k \in 1, \dots, K$, fit an individual model  using {\tt cv.glmnet} and the {\tt  offset} and {\tt penalty.factor}. Use these models for prediction within each group.
\end{enumerate}
\end{algorithm}

We again note that using $\alpha=0$ is similar to using the overall model for each class: it uses the same support set, but ``fine-tunes'' the weights (coefficients) to better fit the specific group. When $\alpha=1$ the method corresponds to fitting $k$ separate class-specific models. 
\medskip

\begin{remark}
\normalfont   
\small
    The forms for the offset and penalty factor were chosen so that the family of models, indexed by $\alpha$, captures both the individual and overall models at the extremes. We have not proven that this 
    formulation is optimal in any sense, and a better form may exist.
\end{remark}

  \begin{remark}  
  \normalfont
  \small
    We can think of the pretrained lasso as a simple form of a Bayes procedure, in which we pass ``prior'' information --- the offset and penalty factor --- from the first stage model
    to the individual models at the second stage.
  \end{remark}

\subsection{Example: TCGA pancancer dataset}
\begin{figure}
\centering
\includegraphics[width=.6 \linewidth]{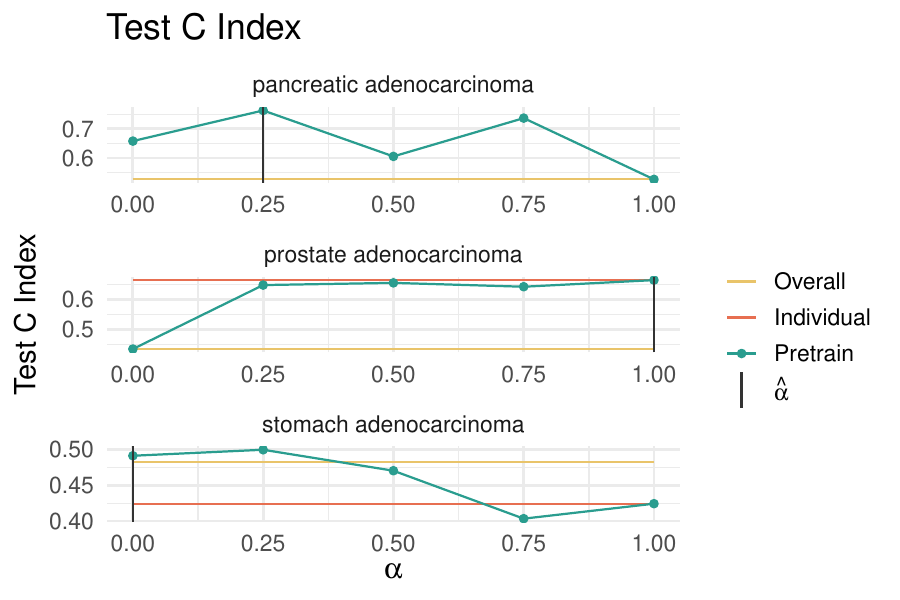}
\caption{\em TCGA dataset: C-index values for different models across 3 cancer classes. The vertical lines indicate the value of $\hat \alpha$ chosen by cross-validation for each class.}
\label{fig:TCGA}
\end{figure}

We applied pretraining to the public domain TCGA pancancer dataset~\citep{goldman2020visualizing}. After cleaning and collating the data, we were left with 714 patients and 20,531 gene expression values.
The patients fell into one of 3 cancer classes as detailed in Table \ref{tab:tab1}. The outcome was DFI (disease-free interval): there were 160 events.
For computational ease, we filtered the genes down to the 2000 genes having the largest absolute Cox PH score, and all methods used this filtered dataset.
The data was divided into a training and testing sets with an $80\%/20\%$ split. We used cross-validation to select $\alpha$ for each group. Results are in Figure~\ref{fig:TCGA}.

\begin{table}[ht]
\centering
\caption{\label{tab:tab1} \em Sample size, support size and test C-index for each fitted model. ``Total C-index'' refers to the C-index computed using the full dataset. Pretraining results use the cross-validated choice of $\alpha$ for each class, as in Figure~\ref{fig:TCGA}. Pretraining outperforms the overall and individual models, and maintains a small support size.}
\begin{small}
\begin{tabular}{r|rrrr}
  \hline
 & Pancreatic & Prostate & Stomach & Total \\ 
  \hline
\textbf{Sample size} & 72 & 384   & 258 & 714 \\ 
 \hline
 \textbf{Support size} & & & & \\
  Overall & -- & -- & -- &  16 \\ 
  Pretrain  & 22 & 27 & 16 & 33\\ 
  Individual & 1 & 11 & 40 & 52\\ 
   \hline
   \textbf{Test C-Index} & & & & \\
  Overall & 0.52 & 0.43 & 0.48 &  0.62 \\ 
  Pretrain  & \textbf{0.76} & \textbf{0.67} & \textbf{0.49} & \textbf{0.69}\\ 
  Individual & 0.53 & \textbf{0.67} & 0.42 & 0.63\\ 
  \hline
\end{tabular}
\end{small}
\end{table}

\subsection{Pretraining using an external dataset}
\label{sec:external}
\begin{figure}
    \centering
\includegraphics[width=.9 \linewidth]{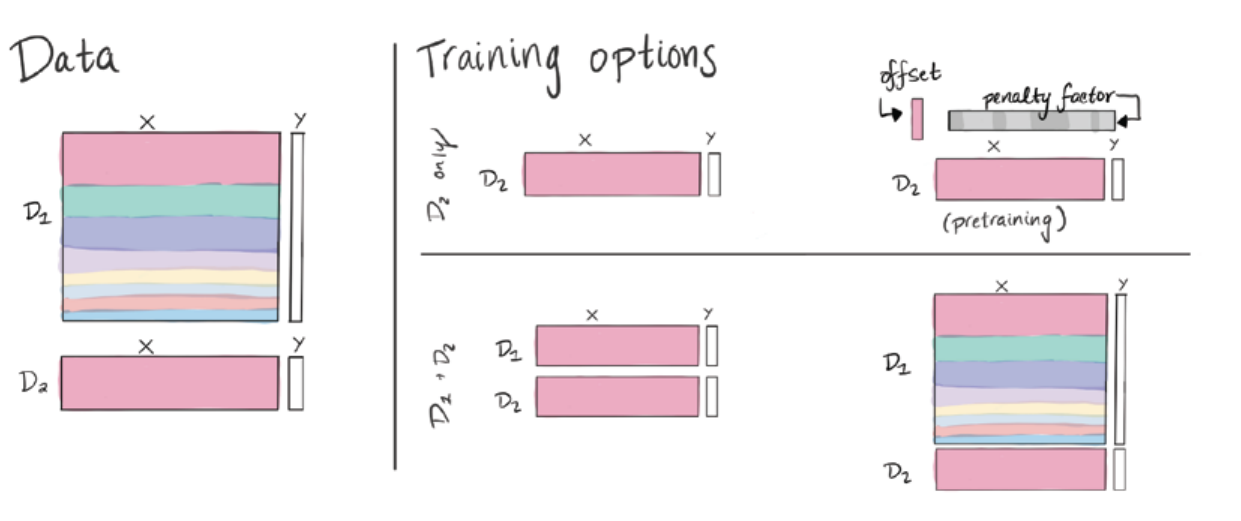}
\caption{\em Options for modeling with an external dataset $D_1$ and a smaller training dataset $D_2$. (The bottom row depicts options when $D_1$ is available at train time.)}
\label{fig:externalSchem}
\end{figure}

Here we examine the setting where there is a large external dataset with multiple classes (denoted by $D_1$),
and we have a smaller training set  ($D_2$) from just one class (say class 1). Our goal is to make accurate predictions
for class 1. This follows our analogy to using ImageNet ($D_1$) to train a large neural network, and then to fine tune using a smaller dataset of cats and dogs ($D_2$). We consider four different approaches to this problem: (1) Fit the lasso with cross-validation ({\tt cv.glmnet}) to $D_2$;
(2) Run the pretrained lasso, using $D_1$, $D_2$ for the two stages;
(3) Combine the data from $D_1$ and $D_2$ for class 1, run {\tt cv.glmnet} on this class 1 data; (4) Combine  all of $D_1$ with $D_2$, run {\tt cv.glmnet}. 
This is illustrated in Figure~\ref{fig:externalSchem}.

Note that (1) does not require access to $D_1$, while (2) requires just the offset and penalty factor from the lasso
model fit to $D_1$. On the other hand, (3) uses class 1 data from $D_1$, while (4) requires {\em all} of the data $D_1$.
We wish to compare approaches (1) and (2), which do not require access to $D_1$, with the other two approaches.

We conducted a simulation study to evaluate these four approaches. We simulated a 10-group dataset $D_1$ with 100 samples per group and 500 features, and an additional dataset $D_2$ with 100 samples drawn from the same distribution as group 1 in $D_1$. For each group $g$, the coefficients $\beta_g$ were generated as a mixture of shared coefficients and group-specific coefficients, with blending controlled by a parameter $\rho$: 
\[\beta_g = \rho \cdot \beta_{\text{shared}} + (1 - \rho) \cdot \beta_{g,  \text{ indiv}},\]  
where $\beta_{\text{shared}}$ is identical across groups and $\beta_{g,  \text{ indiv}}$ is group specific. The supports for $\beta_{g, \text{ indiv}}$ are non-overlapping across groups. When $\rho = 1$, all groups share the same support and coefficient values; when $\rho = 0$, each group has distinct support (no shared effects). Intermediate values of $\rho$ reflect a mixture of shared and individual effects.

For each simulation, we fit models using all four approaches, and compared performances using prediction squared error (PSE) on a held-out test set. Even when there is no shared support, pretraining using all of $D_1$ did not hurt. When there is shared support, pretraining performs nearly as well as having extra data from $D_1$. Performance for $\rho = 0, 0.5, 1$ is shown in Figure~\ref{fig:external}. In Appendix~\ref{app:simDataExternal}, we repeat this experiment in the setting where the groups share \emph{support}, but the effect sizes are different across groups.

\begin{figure}
\centering
\includegraphics[width=.9\linewidth]{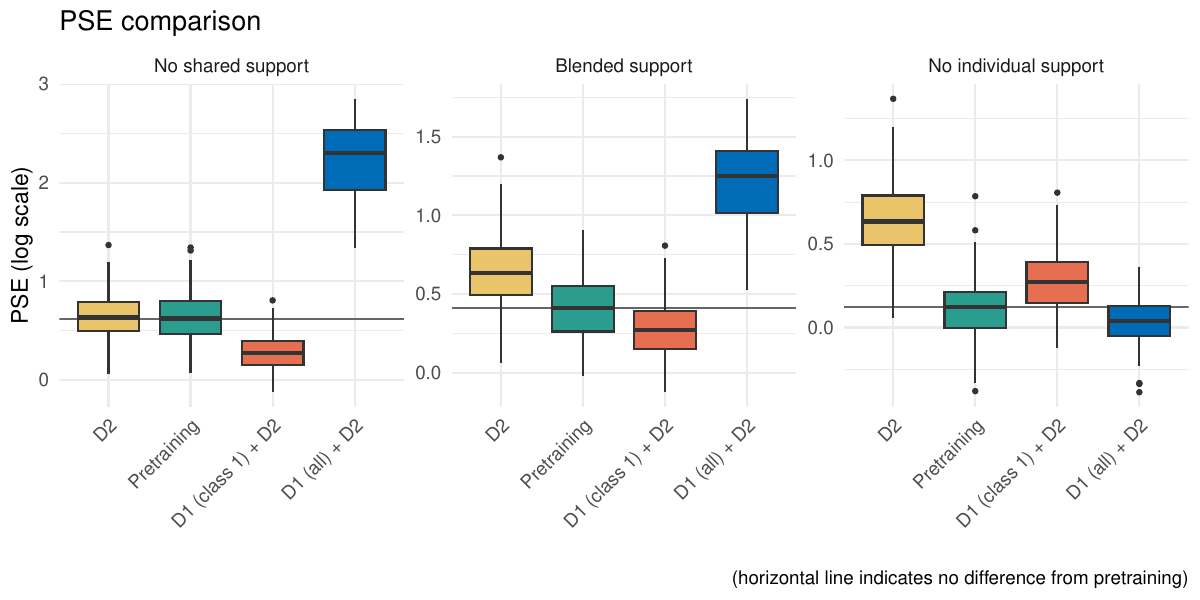}
\caption{\em Comparison of approaches for modeling with a large (usually inaccessible) dataset $D_1$ and a smaller dataset $D_2$. When $D_1$ is informative, pretraining using only $D_2$ and the coefficients from a model trained with $D_1$ performs nearly as well as having access to $D_1$ directly. When $D_1$ is not informative, pretraining with $D_1$ does not hurt.}
\label{fig:external}
\end{figure}

\section{Related work}
\label{sec:related}

\subsection{Data shared lasso}
The data shared lasso~\citep{gross2016data} (DSL) is a closely related approach for modeling data with a natural group structure. It solves the problem
\begin{equation}
\left(\hat{\beta}, \hat{\Delta}_1, \dots, \hat{\Delta}_K\right) = \argmin \frac{1}{2} \sum_{i} \left(y_i - x_i^T (\beta + \Delta_{k_i})\right)^2 + \lambda \left(\|\beta\|_1 + \sum_{k=1}^K r_k \|\Delta_k\|_1\right).
\end{equation}
It jointly fits an overall coefficient vector $\beta$ that is common across all $K$ groups, as well as a modifier vector $\Delta_k$ for each group $k$. The parameter $r_k$ in the penalty term controls the size of $\|\Delta_k\|_1$, and therefore determines whether the solution should be closer to the overall model $\beta$ (for $r_k$ large) or the individual model $\beta + \Delta_k$ (for $r_k$ small). 

DSL is analogous to pretrained lasso in many ways. Both approaches fit overall and individual  models, and both have a parameter ($r_k$ or $\alpha_k$) to balance between the two. One important difference between DSL and pretraining is the use of \texttt{penalty.factor} in pretraining. For $0 < \alpha < 1$, pretraining encourages the individual models to use the same features that are used by the overall model, but allows them to have different values. DSL has no such restriction relating $\beta$ to the modifier $\Delta$. Additionally, because pretraining is performed in two steps, it is flexible: researchers with large datasets can train and share overall models that others can use to train an individual model with a smaller dataset.

\subsection{Laplacian regularized stratified models}
Stratified modeling fits a separate model for each group. \textit{Laplacian regularized} stratified modeling~\citep{tuck2021fitting} incorporates regularization to encourage separate group models to be similar to one another, depending on a user-defined structure indicating similarity between groups. For example, we may expect lymphoma and leukemia to have similar models because they are both blood cancers, and we could pre-specify this when model fitting. While pretrained lasso uses information from an \textit{overall} model, laplacian regularized stratified modeling uses similarities \textit{between groups}. 

\subsection{Reluctant interaction modeling}
Reluctant interaction modeling~\citep{yu2019reluctant} is a method for training a lasso model using both main and interaction effects, while (1) prioritizing main effects and (2) avoiding the computational challenge of training a model using all $p^2$ interaction terms. It uses three steps: in the first, a model is trained using main effects only. Then, a subset of interaction terms are selected based on their correlation with the \emph{residual} from the first model; the intention is to only consider interactions that may explain the remaining signal. Finally, a model is fit to the residual using the main effects and the selected set of interaction effects. Though it has a different goal than pretraining, Reluctant Interaction Modeling shares high level
properties with our current proposal.

\subsection{Mixed effects models}
Mixed effects models jointly find \textit{fixed} effects (common to all the data) and \textit{random} effects (specific to individual instances). A linear mixed effect model has the form 
\begin{equation}
    y = X \beta + Z \theta + \varepsilon,
\end{equation}
where $X$ consists of features shared by all instances and $Z$ consists of features related to individual instances. Both pretrained lasso and mixed effects modeling aim to uncover two components; in pretraining, however $X = Z$ and we seek to divide $\beta$ into overall and group-specific components.

\section{Pretrained lasso: a wide variety of use cases}
\label{sec:useCases}
We have described the main idea for lasso pretraining, as applied to data with grouped observations: a model is fit on a large set of data, an offset and penalty factor are computed, and these components are passed on to a second stage, where individual models are built for each group.
It turns out that the pretraining idea for the lasso is a general paradigm: it is a method for passing information from one model to another, and there are many different ways that it can be applied. Typically the pipeline has only two steps, as in the example above; but in some cases it can consist of multiple steps, as made clear next. Below is a (non-exhaustive) list of potential use cases, {\em the common feature of which is the passing of an offset and penalty factor from one model to the next.}

\begin{enumerate}
\item {\bf Input grouping:}\\The rows of $X$ are partitioned into groups. These groups may be:
\begin{description}
\item{(a)} Pre-specified (Section~\ref{sec:alg}), e.g. cancer classes, age groups, ancestry groups.
The pancancer dataset described above is an example of this use case.
\item{(b)} Pre-specified but different in training and test sets (Section~\ref{sec:diff_groups}), e.g. different train and test patients.
\item{(c)} Learned from the data via a decision tree (Section~\ref{sec:learned_groups}).
\end{description}
\item{\bf Multinomial response:} Predict class membership with $>2$ classes, e.g. predict a disease subtype or a cell type (Section~\ref{sec:multinomial_response}). We wish to leverage shared signal across classes, and to flexibly identify signal specific to each class. 
\item{\bf Multi-response:} Predict a matrix of responses. There are two special cases: time-ordered columns, where the same target is measured at different points in time, and mixed targets, where the target columns are of different types, e.g. quantitative, survival, or binary/multinomial. 
This is illustrated in Sections \ref{sec:multichain} and \ref{sec:timechain}.
\item {\bf Conditional average treatment effect estimation}. This is similar to the input grouping case; groups are defined by the levels of a treatment variable (Section~\ref{sec:CATE}).
\item {\bf Gradient boosting:} We fit a gradient boosting model, and use the individual trees as features in a pretraining pipeline (Section~\ref{sec:boost}). 
\end{enumerate}
Of course, other scenarios are possible. 

\subsection{Multinomial response models}
\label{sec:multinomial_response}
Suppose now that we have no grouping on the rows of $X$; instead we have $K$ response classes and wish to fit a multinomial model. 
 It is much the same as our earlier algorithm, the only difference is the way in which the models are combined at the end. Algorithm~\ref{alg:alg2} describes the procedure in detail, and this is applied to real data in Section~\ref{sec:celltypes}. 

\begin{algorithm}[h]
\caption{Pretrained lasso for multinomial responses}
\label{alg:alg2}
\begin{enumerate}
\item At the first stage, let $B_{p\times K}$  be the coefficient matrix. Fit a grouped multinomial model to all classes: use two-norm penalties on the rows of $B$ (i.e. $\sum_{j=1}^p \|\beta_{j, \cdot}\|_2$).
\item Second stage: for each class $k$, define the offset equal to the  $k$th column of ${(1-\alpha) X\hat B}$. Define $S_k$ to be  the support of the $k$th column of $\hat B$.
Use penalty factor $I(j\in S_k) + \frac{1}{\alpha} \cdot I(j\notin S_k)$. Fit a two class model for class $k$ vs the rest using the offset and penalty factor.
\item Classify each observation to the class having the maximum probability across all of the  one versus rest problems.
\end{enumerate}
\end{algorithm}
\noindent

\subsubsection{Multinomial response example: classifying cell types}
\label{sec:celltypes}

We applied the pretrained lasso together with SPLASH~\citep{CHAUNG20235440, kokot2023splash2}, a new approach to analyzing genomics sequencing data.  SPLASH is a statistics-first alignment-free inferential approach for genomic sequence analysis. SPLASH is applied directly to raw sequencing reads and returns k-mers which show statistical variation across samples.  Here we used the output of SPLASH run on 10x muscle cells (2,760 cells from the 10 most common muscle cell types in donor 1)  from the Tabula Sapiens consortium \citep{the2022tabula}, a comprehensive human single-cell atlas. SPLASH yielded about 800,000 (sparse) features.

\begin{figure}[ht]
    \centering
    \includegraphics[width=.8\textwidth]{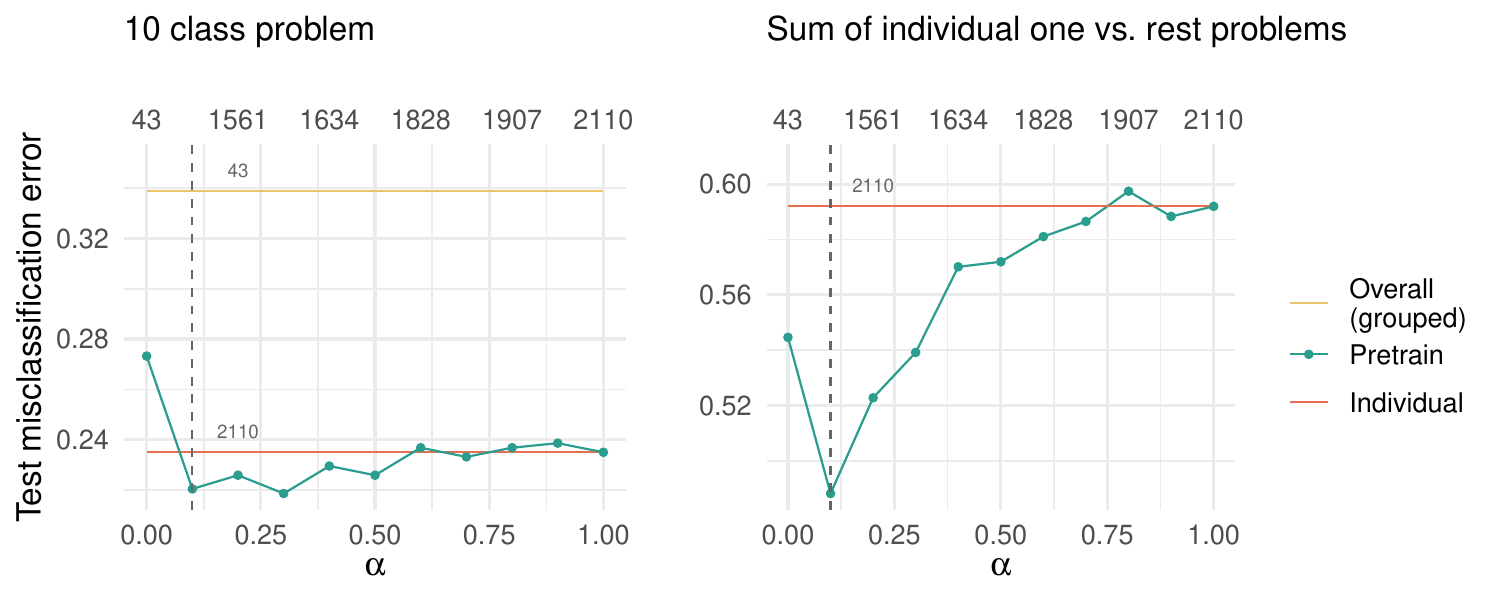}
    \caption{\em Left: misclassification rate of the 10 class problem. Right: the sum of misclassification rates across 10 one-vs-rest problems. The vertical dashed line shows the value of the hyperparameter $\alpha$ chosen by cross-validation.}
    \label{fig:cell_classification}
\end{figure}

 We divided the data into 80\% train and 20\% test sets so that the distribution across the 10 cell types was roughly the same in training and testing, and we used cross-validation to select the pretraining hyperparameter $\alpha$. Figure~\ref{fig:cell_classification} shows results across a range of $\alpha$ values on held-out data. We measure performance with the test data in two ways: (1) we compute the misclassification rate by treating this as a 10 class classification task and (2) we compute the \textit{sum} of misclassification rates by treating this as a set of 10 one vs. rest problems. Using the $\hat{\alpha}$ selected by cross-validation, pretraining outperforms the overall and individual models by both metrics. As a single 10-class problem, the misclassification rate is 0.220 vs 0.339 or 0.235 for the overall and individual models, and it uses fewer selected features than the individual models (1659 features vs 2110 features). 

An important open biological question is to determine which of the features selected by SPLASH are cell-type-specific or predictive of cell type. We tested whether the pretrained lasso could be used to determine which alternative splicing events found by SPLASH were predictive of cell type.  Without tuning, the pretrained lasso reidentified cell-type-specific alternative splicing in MYL6, RPS24, and TPM2, all genes with established cell-type-specific alternative splicing~\citep{olivieri2021rna}. In addition, SPLASH and the pretrained lasso identified a regulated alternative splicing event in Troponin T (TNNT3) in Stromal fast muscles cells which to our knowledge has not been reported before, though it is known to exhibit functionally important splicing regulation~\citep{schilder2012cell}. These results support the precision of SPLASH coupled with the pretrained lasso for single cell alternative splicing analysis.

\subsection{Multi-response models}
\label{sec:multichain}

Another interesting use case is the multi-response setting, where the outcome $Y$ has $K>1$ columns, and the data in these columns may be quantitative or integers.
The multinomial response discussed earlier can be expressed as a multi-response problem corresponding to one-hot encoding of the classes. But the multi-response setup is more general, and can be used in problems where  each observation can fall in more than one class.

To apply the pretrained lasso here, we fit a grouped multi-response model (Gaussian or multinomial) to all of the columns, and then fit individual models to each column separately. In the Gaussian case, the first step uses the grouped multi-response loss:
\begin{equation}
    \frac{1}{2}\sum_{k=1}^K \|y_k - X \beta_{\cdot, k}\|_2^2 + \lambda \sum_{j = 1}^p \|\beta_{j, \cdot}\|_2,
\end{equation}
where $y_k$ is the $k^\text{th}$ response and $\beta_{\cdot, k}$ are the corresponding coefficients. For a particular feature $j$, the penalty $\|\beta_{j, \cdot}\|_2$ forces $\beta_{j, k}$ to be zero or nonzero for all $k = 1, \dots, K$. The second step uses pretraining as usual for each response: the penalty factor and offset for the $k^\text{th}$ response are defined as in Algorithm~\ref{alg:alg1} using the coefficients $\beta_{\cdot, k}$.

\subsubsection{Multi-response example: chemometric data}
\begin{figure}[!h]
    \centering
    \includegraphics[width=.8\linewidth]{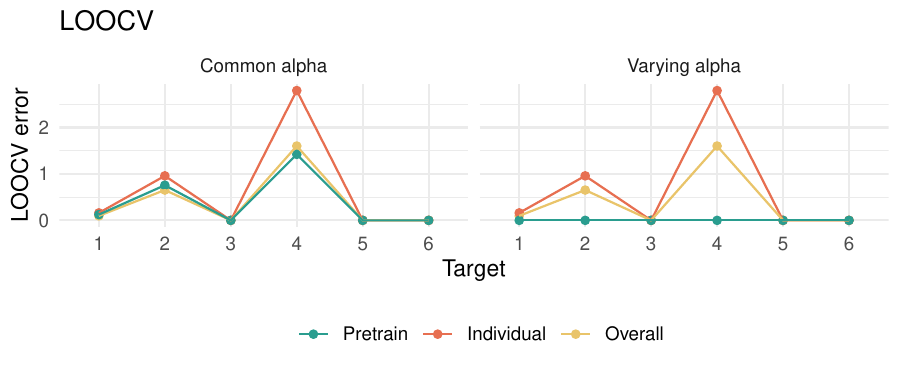}
\caption{\em Results for multi-response chemometrics example.}
\label{fig:chemo}
\end{figure}

Figure \ref{fig:chemo} shows an example taken from \cite{skagerberg1992multivariate}, simulating the production of low-density polyethylene. The data were generated to show that quality control could be performed using measurements taking during polyethylene production to predict properties of the final polymer. The authors simulated 56 samples with 22 features including temperature and solvent flow-rate, and 6 outcomes: number-average molecular weight, weight-average molecular weight, frequency of short chain branching, the content of vinyl groups and vinylidene groups. Figure~\ref{fig:chemo} shows the LOOCV squared error over the 6 outcomes, using
both the best common $\alpha$ (left) and outcome-specific
$\alpha$ values (right). Pretrained lasso
performs best in both settings.

\subsection{Time ordered responses and chaining outcomes}
\label{sec:timechain}
Other multi-response settings include prediction of time-ordered responses, and outcomes of different types (e.g. quantitative, survival, binary).
In both cases we apply pretrained lasso in a sequential fashion.
We fit a model to the first outcome, compute the offset and penalty factor,
and pass these to a model for the second outcome, and so on. This application requires a prior ordering
of the outcomes. In real applications, when an ordering is not already defined,
it might make sense to place the primary outcome measure in the first position,
and the rest in decreasing order of 
importance.

\subsubsection{Time ordered response example: mass cytometry cancer data}
\begin{figure}[!h]
    \centering
    \includegraphics[width=.7\textwidth]{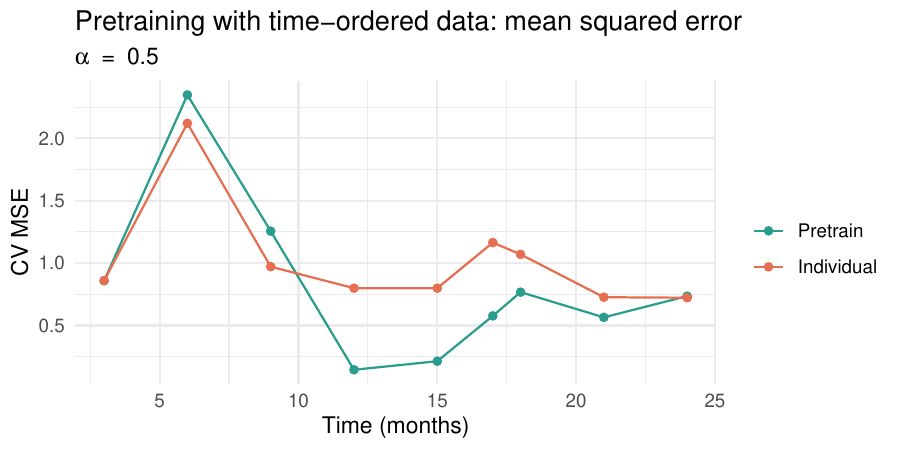}
\caption{\em Results for the time ordered data example. Cross-validation was used to select the pretraining hyperparameter $\alpha = 0.5$.}
\label{fig:nilotinib}
\end{figure}

We applied pretraining to a dataset of mass cytometry measurements from 8 patients with chronic myeloid leukemia who are treated with the drug nilotinib~\citep{gullaksen2017single}. Our objective is to predict each patient’s $\text{BCR::ABL1}^{\text{IS}}$ levels at 9 time points over 24 months, as this highly sensitive measure of residual disease is strongly correlated with progression-free survival. Figure~\ref{fig:nilotinib} shows the results of pretraining applied to this dataset as described for time-ordered responses. For all time points, pretraining nearly matches or outperforms the alternative of fitting separate models. 

\begin{remark} Pretrained lasso fits an interaction model.
\small
{\em In general, suppose we have a response $y$, features $x$  and grouping variables $G_1, G_2, \ldots G_g$.
The grouping variables can stratify the inputs or the target (either multinomial
or multi-response). Introduction of a grouping variable $G_j$ corresponds to the addition of an interaction
term between $x$ and $G_j$. Thus one could imagine a more general forward stepwise pretraining process as follows:

\begin{enumerate}
\item Start with an overall model, predicting $y$ from $x$, without any consideration of the grouping variables.
Let $O_1$ and $\textit{pf}_1$ be the offset and penalty factor from the chosen model.
\item Introduce the grouping variable $G_1$ by fitting individual models to the levels of $G_1$, with
the offset and penalty factor  $O_1$ and $\textit{pf}_1$. From these models extract $O_{2k}$ and $\textit{pf}_{2k}$ for the $K_1$
levels of $G_1$: $k=1,2,\ldots K_1$. 
\item Introduce the grouping variable $G_2$, either as an interaction $x\times G_2$ or an interaction
$x\times G_1 \times G_2$, and so on.
\end{enumerate}
}
\end{remark}

\subsection{Different groupings in the train and test data}
\label{sec:diff_groups}
 In our initial ``input grouped'' setting, our training data are partitioned into groups, and we observe the same groups at test time. Now, we consider the setting where the test groups were not observed at train time. For example, we may have a training set of \emph{people}, each of whom has many observations, and at test time we wish to make predictions for observations from new people.

\begin{figure}[!ht]
    \centering
    \includegraphics[width=.95 \textwidth]{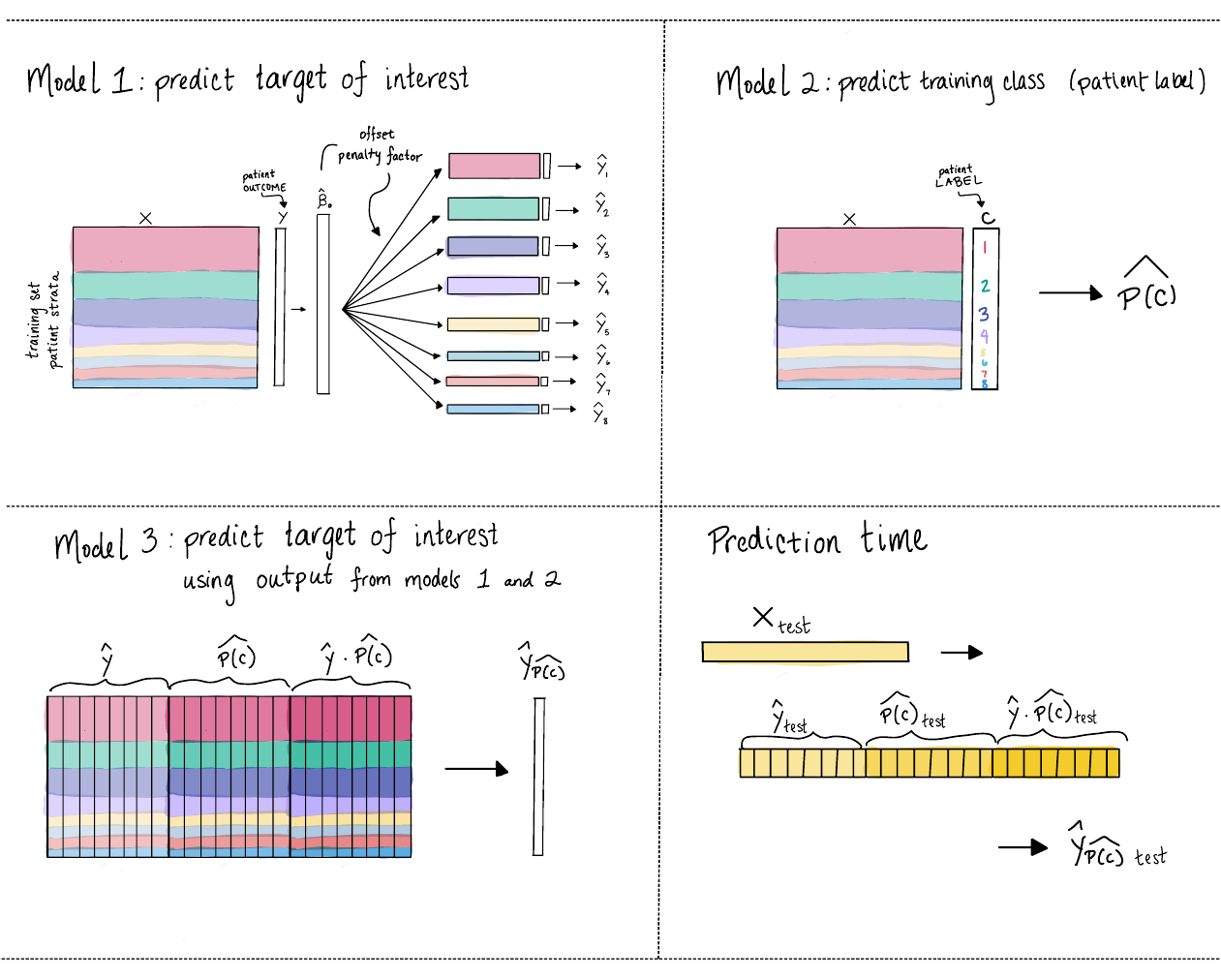}
    \caption{\em How to use pretraining when the train and test strata are different (Section~\ref{sec:diff_groups}).}
    \label{fig:by_patient}
\end{figure}

 To address this, we use pretraining as described. Now, however, we fit an extra classifier 
 to predict the \textit{training group} for each observation. 
We then train a final model that predicts the response using the output from the pretraining models and the group classifier. 
 The procedure is illustrated in Figure~\ref{fig:by_patient}, 
 and applied to real data in Section~\ref{sec:massspec}.

Although one could use the estimated posterior class probabilities from this construction, we have found better empirical results by training a supervised learning algorithm to predict
$r(x_i)$ from $\hat p_k(x_i), \hat q_k(x_i)$ and $\hat p_k(x_i)\cdot \hat q_k(x_i), k=1,2,\ldots K$.

\subsubsection{Different train/test grouping example: mass spectrometry cancer data}
\label{sec:massspec}

This data, from a melanoma proteomics study~\citep{margulis2018distinguishing}, has 2094 peak heights measured via DESI mass spectrometry per pixel, with about 1,000 pixels from each patient.
 There are  28 training patients, 15 test patients; a total of 29,107 training pixels and 20,607 test pixels. The output target is binary (healthy vs disease). All error rates quoted are per pixel rates.

We clustered the training patients using K-means into 4 groups, and then applied pretraining. The pretrained lasso provides an advantage in AUC relative to the overall model: $0.96$ for pretraining vs $0.94$ without.

\subsection{Learning the input groups}
\label{sec:learned_groups}
Here we consider the setting where there are no fixed input groups, but instead we learn potentially useful input groups from a CART decision tree, and then apply pretraining with these groups. Typically, the  features that we use for splitting are not the full set of features $x$ but instead a small set of clinical variables that are meaningful
to the scientist.

\subsubsection{Learning the input groups example: predicting myocardial infarction}
We illustrate this on the U.K. Biobank data, where we have derived 299 features on 64,722 white British individuals, and we wish to predict myocardial infarction.  There are 249 metabolites from  nuclear magnetic resonance and 50 genomic PCs. The features available for splitting were age, PRS (polygenic risk score) and sex (0=female, 1=male). 

\begin{figure}
    \centering
    \begin{subfigure}[h]{.45 \textwidth}
    \centering
    \includegraphics[width=\linewidth]{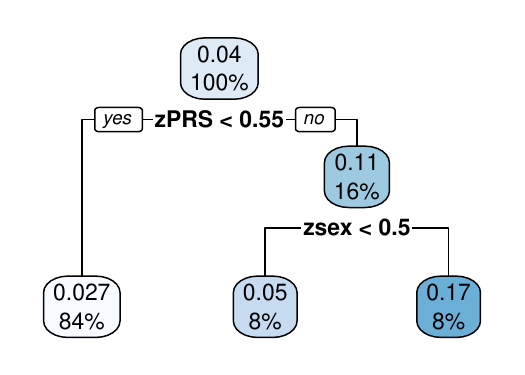}
    \end{subfigure}
    \hfill
    \begin{subfigure}[h]{.5\textwidth}
    \includegraphics[width=\linewidth]{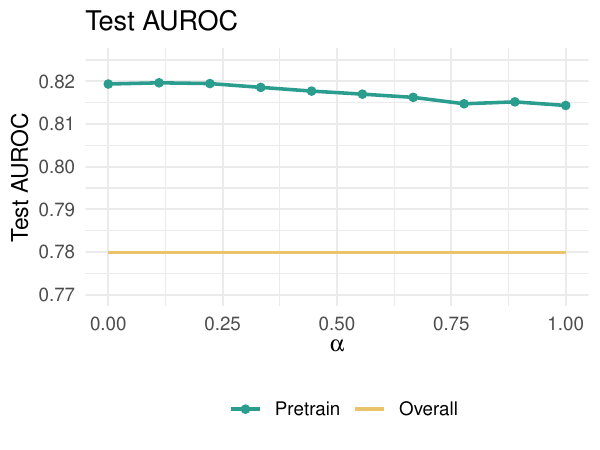}
    \end{subfigure}
    \caption{\em Left: CART tree learned from the features PRS, Sex and Age.  Right: test set AUC for overall (no groups) model and pretrained lasso applied to the 3 groups.}
    \label{fig:learned}
\end{figure}

We used a 50/50 train/test split and built a CART tree using the R package {\tt rpart}, limiting the depth of the tree to be 3
(for illustration)\footnote{Another way to grow the decision tree in this procedure  would be to use  ``Oblique Decision Trees'', implemented in the ODRF R language package. These trees fit linear combinations of the features at each split. Since the pretrained lasso fits a linear model (rather than a constant) in each terminal node, this seems natural here.
We tried ODT in this example: it produced a similar tree to that from CART, and hence we omit the details. We thank Yu Liu and Yingcun Xia for implementing changes to their R package ODRF so that we could use it in our setting.}. The left panel of Figure \ref{fig:learned} shows the resulting tree.
The right-most terminal node contains men with high PRS scores: their risk of MI is much higher than the other two groups (0.17 versus 0.027 and 0.05). The predictions using just this CART tree had a test AUC of 0.49.

We then applied the pretrained lasso for fixed input
groups (Algorithm \ref{alg:alg1}) to the three groups
defined by the terminal nodes of the tree.
The resulting test AUCs for the pretrained lasso and the
overall model (an $\ell_1$-regularized logistic regression) are shown in the right panel. The pretrained lasso delivered about a 4-5\% AUC advantage for all values of $\alpha$, and it identified a strong interaction of one feature with PRS and sex.

\subsection{Conditional average treatment effect estimation}
\label{sec:CATE}
\def\argmin{\rm argmin}
An important problem in causal inference is estimation of the conditional average treatment effect (CATE). The data is of the form $ (X_i, W_i, y_i), i=1,2,\ldots n$ where $X_i$ is a vector
of covariates, $Y_i$ is a quantitative outcome and $W_i$ is a binary treatment indicator. We denote by 
$(Y_i(0),Y_i(1))$ the corresponding outcomes we would have observed given the treatment assignment $W_i=0$ or 1 respectively.
The goal is to estimate the CATE:  $\tau(x)\equiv E(Y(1)|x)-E(Y(0)|x)$. We make the usual assumption that the treatment assignment is unconfounded.

One popular approach is the ``R-learner'' of \cite{nie2021quasi}. It is based on the objective function
\begin{equation}
\hat\tau=\arg \min_\tau \frac{1}{n}\sum\Bigl[ (Y_i-m(X_i))-(W_i-e(X_i))\cdot\tau(X_i) \Bigr]^2
\label{eqn:rlearner}
\end{equation}
where $m(x)$ is the overall mean function and $e(x)$ is the treatment propensity ${{\rm Pr}( W=1| X=x)}$.
In the simplest case (which we focus on here), a linear model is used for $m(x)$ and $\tau(x)$.
The lasso version of the R-learner adds an $\ell_1$ penalty to the objective function. 

The steps of the R-learner are as follows: (1) estimate $m(\cdot),  e(\cdot)$ by fitting $Y$ on $X$, $W$ on $X$, using cross-fitting; (2) estimate $\tau(\cdot)$ by solving (\ref{eqn:rlearner}) above. For simplicity we assume here that the treatment is randomized so that we can set $e(x)=0.5$.

Now we can combine the R-learner with the pretrained lasso as follows. We assume the shared support model 
\begin{eqnarray}
\hat Y=\hat \beta_0+X\hat \beta+W\cdot \hat \tau(X) ;\;\;\;\hat \tau(X)= X\hat \theta_0+X\hat \theta
\label{eqn:rlearn}
\end{eqnarray}
where $\theta$ has the same support and signs as $\beta$.
To fit this, we use the R-learner procedure  above, but include in the  model for $\tau(X)$ the penalty factor computed from the model for $Y$ [we do not include the offset, since the target
in the two models are different].
If this shared support assumption is true or approximately true, we can potentially do a better job at estimating $\tau(x)$. This assumption seems reasonable: it says that the predictive features are likely to overlap with the features that modify the treatment effect.

\begin{figure}
    \centering
    \includegraphics[width=.48\textwidth]{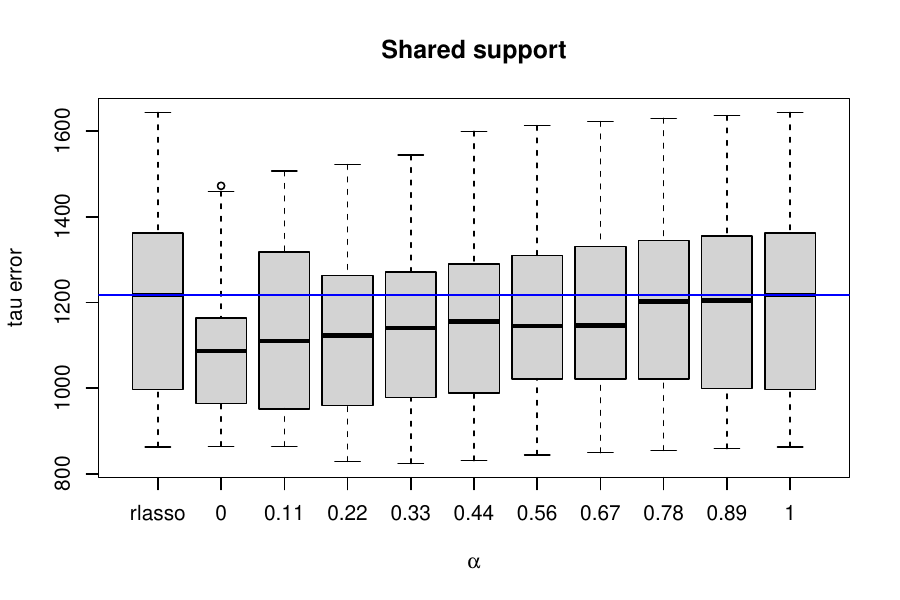}
    \includegraphics[width=.48\textwidth]{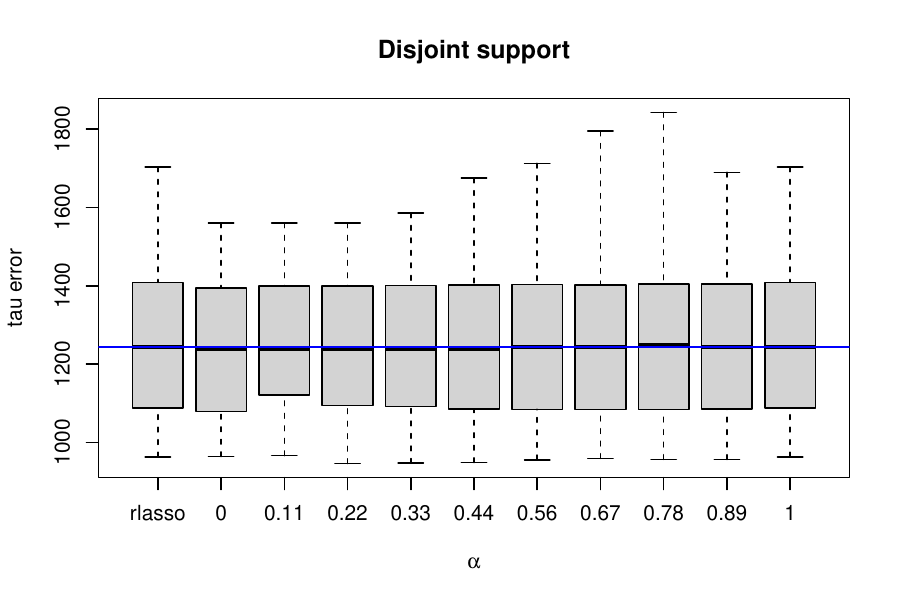}
          \vskip -.3in
\caption{\em Results for R-learner experiment. Horizontal blue line shows the median error for the rlasso R-learner. }
\label{fig:rlasso}
\end{figure}

Figure \ref{fig:rlasso} shows an example with $n=300$, $p=20$ and an SNR of about 2. The first 10 components of $\beta$ are positive, while the second 10 components are zero. In left panel 
the treatment effect $\theta$ has the same support and signs as $\beta$, while in the right panel, its support is in the second 10 features, with no overlap with the support of $\beta$.  The figure shows boxplots of the absolute estimation error in $\tau(x)$ over 20 realizations. 

In the left panel we see that the pretrained R-learner outperforms the R-learner for all $\alpha$, while in the right panel, they behave very similarly. It seems that there is little downside in assuming the shared support model. Upon closer examination, the reason 
becomes clear: under Model~(\ref{eqn:rlearn}) with disjoint support,
all 20 features are predictive of the outcome,
and hence there is no support restriction resulting 
from the outcome model.

\subsection{Beyond linear models: gradient boosting}
\label{sec:boost}
Here we explore the use of basis functions beyond the linear functions used throughout the paper.
Suppose we run gradient boosting~\citep{friedman2000}
for $M$ steps, giving $M$ trees.
Then we can consider the evaluated trees as our new variables, yielding a
new set of features.
We then apply the pretrained lasso to
these new features. Here is the procedure in a little more detail: (1) run $M$ iterations of xgboost to get $M$  trees (basis functions) $B$ ($n\times M$) and (2) run the pretrained lasso on $B$.

Consider this procedure in the fixed input groupings use case.
We use the lasso to estimate optimal weights for each of the trees, both for an overall model, 
and for individual group models.
For the usual lasso, this kind of ``post-fitting'' is not new (see e.g. RuleFit~\citep{friedman2008predictive}, \citep{hastie2009elements} page 622).

\begin{figure}
    \centering
    \includegraphics[width=.5\textwidth]{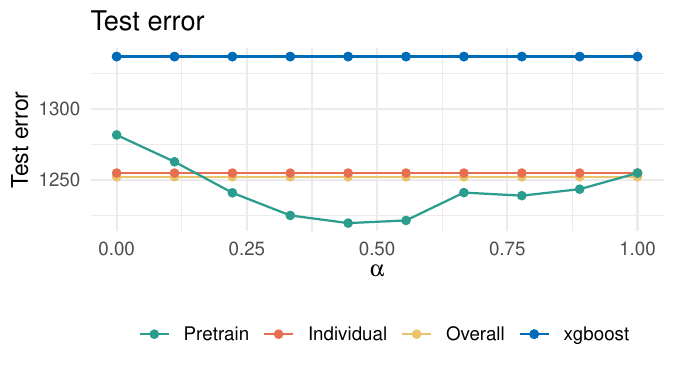}
\caption{\em Results for the pretrained lasso applied to boosted trees.
The first method {\tt xgboost} uses raw features. The remaining use evaluated trees from {\tt xgboost} as features.}
\label{fig:boost}
\end{figure}

It is easy to implement this procedure using the  {\tt xgboost} library in R~\citep{xgboost_r}.
Figure \ref{fig:boost} shows the results from a simulated example. 
We first used {\em xgboost} to  generate
50 trees of depth 1 (stumps). Then  we simulated data using these trees as features, with a strong common weight vector
$\hat\beta_0$. The first method in our example,
{\tt xgboost}, is boosting applied to the raw features, while the other three methods use the 50 trees generated by {\tt xgboost}.
We see that lasso pretraining can help boosting as well.

\section{Theoretical results on support recovery}
\label{sec:theory}
In this section, we prove that the pretraining process recovers the true support and characterize the structure of the learned parameters under suitable assumptions on the training data.
\subsection{Preliminaries}

We call a random variable $Y$ sub-Gaussian if it is centered, i.e., $\mathbb{E}[Y] = 0$ and 
    ${\mathbb{E} \left[ e^{s Y}\right] \le e^{\frac{\sigma^2s^2}{2}}}$ $\quad{\forall s\in\mathbb{R}}$,
where $\sigma^2$ is called the variance proxy of $Y$. For such a random variable, the following tail bound holds:
    $\mathbb{P} \left[ |Y|>t\right ] \le 2 e^{-t^2/(2\sigma^2)}$ for all $t>0$.
The random variable $Y$ has bounded variance when $\sigma^2$ is bounded by a constant. Examples of such random variables include the standard Gaussian and any random variable that is zero-mean and bounded by a constant.

\subsection{An overview of the theoretical results}
\subsubsection{Conditions for deterministic designs}
The recovery of the support of coefficients in lasso models is traditionally guaranteed by conditions like the irrepresentability condition. In this paper, we extend these conditions to the pretrained lasso. Specifically, we introduce a set of deterministic conditions that ensure the recovery of the true support in the shared support model, even when observations are mixed. These conditions, referred to as Pretraining Irrepresentability Conditions, are necessary and sufficient for the pretraining estimator to discard irrelevant variables and recover the true support. Although these conditions are slightly more complex than the classical irrepresentability condition due to the mixed observation model, they are easily interpretable. In summary, there are three key requirements: (1) the off-support features need to be incoherent with the features in the support, (2) the empirical covariance of the features in the support need to be well conditioned, (3) the individual parameters need to be bounded in magnitude.

\subsubsection{Conditions for random designs}
Two key aspects are studied when the design matrix is random:
\begin{itemize}

\item Pretraining under isometric features: We introduce the subgroup isometry condition \eqref{eq:subgroup_isometry} to capture how representative the empirical covariance of a subgroup is in relation to the full dataset. This condition holds when the features are independent sub-Gaussian random variables, and helps in analyzing the behavior of the pretraining estimator.

\item Recovery under sub-Gaussian covariates: We prove in Theorems \ref{thm:recovery_model1} and \ref{thm:exact_recovery_model1} that under certain conditions on the sample size, variable bounds, and noise levels, the pretraining estimator can recover the true support with high probability under the shared support model \eqref{eq:model_shared} where the supports are common. These results are similar in spirit to the existing recovery results for lasso \citep{wainwright2009sharp}, with a few crucial differences. In particular, it is known in the classical setting that $\mathcal{O}(s\log(p-s))$ measurements are necessary for support recovery with high probability. However, in our setting, our result given in Theorem \ref{thm:recovery_model1} show that the   number of measurements needs to scale as $\mathcal{O}(\max(1,\gamma^2) s\log(p-s))$, where $\gamma$ is an upper-bound on the magnitude of the weights $|\beta_k|\,\forall k$. The extra $\max(1,\gamma^2)$ factor is due to the mixture observation model \eqref{eq:model_shared} instead of a simple linear relation studied in earlier literature. It is an open question to verify that this factor is unavoidable, which we leave for future work.

\end{itemize}
In addition, we extend the shared support model \eqref{eq:model_shared} to lift the assumption that the supports are common, and consider the common and individual support model \eqref{eq:model_common_and_indiv}. In this model, there is a shared support between the groups, as well as additional individual supports. We show that support recovery results given in Theorems \ref{thm:recovery_model1} and \ref{thm:exact_recovery_model1} still hold under the assumption that the magnitudes $\beta_k^*$ that belong to the individual support are sufficiently small for each $k$. This is a necessary condition to ensure that the pretraining estimator only recovers the common support and discards individual supports for each group.

\subsection{Shared support model}
Consider $K$ sets of observations 
\begin{align}
y_k=X_k\beta^*_k + \varepsilon_k\in \mathbb{R}^{n/K}, \label{eq:model_shared}
\end{align}
where $\beta^*_k$ are unknown vectors which share a support $S$ of size $s$. More precisely, we have $(\beta^*_k)_{S^c}=0~\forall k\in[K]$ where $S^c$ is the complement of the subset $S$. Here, $\varepsilon=[\varepsilon_1,\hdots,\varepsilon_K]$ is a noise vector to account for the measurement errors, which are initially assumed to be deterministic. We assume that $\frac{n}{K}$ is an integer and each subgroup has at least $s$ samples, i.e., $\frac{n}{K}\ge s$.
We let $X:= \begin{bmatrix} X_1^T&\hdots & X_K^T \end{bmatrix}^T \in \mathbb{R}^{n\times p}$ to denote the full dataset and observations $y:= \begin{bmatrix} y_1&\hdots& y_K \end{bmatrix} \in \mathbb{R}^{n}$.

We define the pretraining estimator as
\begin{align}
\label{eq:pretraining_lasso}
    \hat \beta_{\textrm{pre}} = \arg\min_{\beta} \frac{1}{n}\| y-X\beta\|_2^2 + \lambda \|\beta\|_1.
\end{align}
Next, we provide an analysis of the pretraining estimator under the shared support model.

\subsubsection{Pretraining irrepresentability conditions}
We now provide a set of deterministic conditions that guarantee that the pretraining estimator $\hat \beta_\mathrm{pre}$ recovers the support of $\beta_1,\hdots,\beta_K$. Note that existing results on support recovery for lasso including the irrepresentability condition \citep{zhao2006model}, restricted isometry property \citep{van2009conditions} or random designs \citep{wainwright2009sharp}  are not applicable due to the mixed observation model in \eqref{eq:model_shared}.

Recall that in the shared support model, the vectors $\beta_1^*,\hdots,\beta_K^*$ have the same support $S$ of size $s$. Let us use $X_S\in \mathbb{R}^{n\times s}$ to denote the submatrix of the data matrix $X$ restricted to the support $S$. 
\begin{lemma}[Pretraining irrepresentability]
\label{lem:irrep_cond}
Suppose that the conditions

\begin{align}
    \|X_{S^c}^T X_S^\dagger \signb(\beta^*_S)\|_{\infty} &< \frac{1}{2} \label{eq:irrep1}\\
     \|X_{S^c}^T P_S^\perp \sum_{k=1}^K D_k X\beta_k^* \|_{\infty}&\le \frac{\lambda}{4} \label{eq:irrep2}\,,
\end{align}
hold. Then, the pretraining estimator discards the complement of the true support $S$, i.e., $j\notin S\,\implies (\hat \beta_{\mathrm{pre}})_j=0\,\forall j$ in the noiseless case, i.e., $n=0$. In the noisy case, if the condition
\begin{align}
\|X_{S^c}^T P_S^\perp  \varepsilon\|_{\infty} &\le \frac{  \lambda}{4}
\end{align}
holds, the same result holds for an arbitrary noise vector $n$.
Here, $\signb(\beta^*_S)=\signb((\beta^*_k)_S\,\forall k\in[K]$ is the sign of the vectors $\beta^*_1,\hdots,\beta_K^*$ constrained to their support S, $P_S^\perp := I - X_S (X_S^TX_S)^{-1} X_S^T$ and $D_k$ is the diagonal selector matrix for the $k$-th set of samples, i.e., $D_kX \in\mathbb{R}^{n\times p}$ is $X_k\in\mathbb{R}^{\frac{n}{K}\times p}$ padded with zeros.
\end{lemma}
\begin{remark}
\normalfont
\small
    When $K=1$ and $D_1=I$, the conditions \eqref{eq:irrep1} and \eqref{eq:irrep2} simplify to the well-known strong irrepresentability condition \citep{zhao2006model}, noting that $P_S^\perp D_k X = P_S^\perp X = 0$ which shows $\eqref{eq:irrep2}$ always holds.  
\end{remark}

\subsubsection{Pretraining under isometric features}
We now analyze the behaviour of the solution $\betapre$ under the assumptions that the samples from the subgroups follow an isometric distribution relative to the entire dataset. 

We introduce the subgroup isometry condition:
\begin{align}
    \textrm{(subgroup isometry)} \qquad \qquad \Big\| (X_S^TX_S)^{-1} (X_S^T D_k X_S) - \frac{1}{K} I \Big \|_2 \le \delta~\forall k\in[K], \label{eq:subgroup_isometry}
\end{align}
for some $\delta \in (0,1)$. The above quantity represents the ratio of empirical covariances of features restricted to the subset $S$, comparing the entire dataset with the subgroup defined by the $k$-th group of samples. It quantifies how representative the empirical covariance of the subgroup is in relation to the full dataset.
\begin{remark}
\normalfont
\small
    Note that we have $(X_S^TX_S)^{-1} (X_S^T D_k X_S)=\big( \sum_{i\in[n]} \tilde x_i \tilde x_i^T \big)^{-1}\big (\sum_{i\in \mathcal{G}_k} \tilde x_i \tilde x_i^T  \big)$, where $\mathcal{G}_k$ is the subset of samples that belong to the group $k$ and $\{\tilde x_i\}_{i=1}^n\in\mathbb{R}^{s}$ are features restricted to the true support $S$. 
\end{remark}

\begin{lemma}
\label{lemma:isometry}
    Suppose that the samples $x_1,\hdots, x_n \in \mathbb{R}^p$ are i.i.d. sub-Gaussian variables with bounded variance and let $n\ge C\delta^{-2} s\log K$, where $C$ is a constant. Then, the subgroup isometry condition in \eqref{eq:subgroup_isometry} hold with probability at least $1-C^\prime e^{-C^{\prime\prime} n}$, where $C^\prime$ and $C^{\prime\prime}$ are constants.
\end{lemma}

\begin{lemma}
\label{lemma:approx_avg}
[Pretraining approximates the average of individual parameters]
    Under the subgroup isometry condition \eqref{eq:subgroup_isometry} and the conditions of Lemma \ref{lem:irrep_cond}, the pretraining estimator satisfies
    \begin{align}
       \Big\| \betapre - \frac{1}{K} \sum_{k=1}^K \beta^*_k + \lambda (X_S^TX_S)^{-1} \signb(\beta_S^*)\Big \|_2\le \delta\sum_{k=1}^K \|\beta^*_k\|_2 + \|X_S^\dagger n\|_2.
    \end{align}
\end{lemma}
\begin{remark}
\normalfont
\small
    The above result shows that the pretraining estimator approximates the average of the individual models $\frac{1}{K} \sum_{k=1}^K \beta^*_k$, in addition to a shrinkage term proportional to $\lambda$.
\end{remark}
The above result is derived from the optimality conditions for the lasso model (see Appendix).

\subsubsection{Recovery under random design}
Next, we prove that the Pretraining Irrepresentability condition holds with high probability when the features are generated from a random ensemble.
\begin{theorem}
\label{thm:recovery_model1}
Suppose that the samples $x_1,\hdots, x_n \in \mathbb{R}^p$ are i.i.d. sub-Gaussian variables with bounded variance and the noise vectors $\varepsilon_1,\hdots,\varepsilon_K$ are sub-Gaussian with variance proxy $\sigma^2$. In addition, assume that $|\beta^*_k|\le \gamma\,\forall k\in[K]$ for some $\gamma>0$, and the number of samples satisfy
\begin{align} \label{eq:randomdesign0}
    n\ge C_1 \max(1,\gamma^2)\, s \log (p-s),
\end{align}
for some constant $C_1>0$. Then, the conditions \eqref{eq:irrep1} and \eqref{eq:irrep2} when $\lambda=C_\lambda \sigma \sqrt{\frac{\log(p-s)}{n}}$ hold with probability at least $1-C_3e^{-C_4n/(s\gamma^2)}$ where $C_2,C_3,C_4$ are constants. Therefore, $\hat \beta_{\mathrm{pre}}$ discards the complement of the true support $S$ with the same probability.
\end{theorem}

\begin{remark}
\normalfont
\small
    It is instructive to compare the condition \eqref{eq:randomdesign0} with the known results on recovery with lasso under the classical linear observation setting \citep{wainwright2009sharp} that require $\mathcal{O}(s\log(p-s))$ observations. Therefore, using only the individual observations without pretraining via the ordinary lasso, we need $n/K > s \log(p-s)$ to achieve the same support recovery. Comparing this with Theorem \ref{thm:recovery_model1}, we observe that the pretraining procedure gives a factor $K$ improvement in the required sample size.
\end{remark}
\begin{remark}
\normalfont
\small
    We note that the factor $\max(1,\gamma)$ is the extra cost on the number of samples induced by the mixture observation model, which is due to the second condition \eqref{eq:irrep2}. In order the pretraining estimator to discard irrelevant variables, the magnitude of each linear model weight $\beta_k^*$ is required to be small. Furthermore, the pretraining estimator can discard variables from the true support $S$.
\end{remark}

\begin{theorem}
\label{thm:exact_recovery_model1}
Suppose that the samples $x_1,\hdots, x_n \in \mathbb{R}^p$ are i.i.d. sub-Gaussian variables with bounded variance and the noise vectors $\varepsilon_1,\hdots,\varepsilon_K$ are sub-Gaussian with variance proxy $\sigma^2$. Set $\lambda =C_\lambda \sigma \sqrt{\frac{{\log(p-s)}}{n}}$. In addition, assume that $|(\beta^*_k)_j|\le \gamma\,\forall k\in[K]\,\forall j\in[p]$ for some $\gamma>0$, and the number of samples satisfy
\begin{align} \label{eq:randomdesign1}
    n\ge C_1 \max(1,\gamma^2)\, s \log\big(\max(p-s,K)\big),
\end{align}
and
\begin{align}
    \min_{j} |\big( \frac{1}{K} \sum_{k=1}^K \beta^*_k \big )_j| \ge C_1^\prime \sigma \sqrt{\frac{\log(p-s)}{n}},
    \label{eq:betamin_condition}
\end{align}
for some constants $C_1, C_1^\prime$.
Then, the pretraining estimator $\betapre$ exactly recovers the ground truth support with probability at least $1-C_3e^{-C_4n}$ where $C_\lambda, C_2,C_3,C_4$ are constants.
\end{theorem}
\begin{remark}
\normalfont
\small
    The condition \eqref{eq:betamin_condition} is similar to the $\beta_{\min}$ conditions used in the classical analysis of lasso under the linear setting \citep{wainwright2009sharp}. This condition is unavoidable to make sure the pretraining estimator does not discard variables in the ground truth support. 
\end{remark}
\begin{remark}
\normalfont
\small
    Note that there are pathological cases for which the pretraining estimator fails to recover the support. A simple example is where $\beta_2^*=-\beta_1^*$ and $K=2$. In this case, the condition \eqref{eq:betamin_condition} can not hold since $\frac{1}{2} (\beta_1^*+\beta_2^*)= 0$. However, we are guaranteed that the pretraining estimator discards all the variables not in the ground truth support under the remaining assumptions.
\end{remark}

\subsection{Common and individual support model}
We now consider $K$ sets of observations 
\begin{align}
y_k=X_k \beta^*_0+X_k\beta^*_k  + \varepsilon_k \in \mathbb{R}^{n/K}, \label{eq:model_common_and_indiv}
\end{align}
where $\beta^*_0$ is an $s$ sparse vector to account for  shared features and $\beta^*_k$ are $s$ sparse unknown vectors modeling individual features. We assume that the support of $\beta^*_0$ and $\beta^*_k$ are not-overlapping.

\subsubsection{Recovery under random design}
Next, we prove that the pretraining estimator discards the complement of the true support with high probability when the features are generated from a random ensemble and the magnitude of the individual coefficients are sufficiently small.
\begin{theorem}
\label{thm:recovery_model2}
Suppose that the samples $x_1,\hdots, x_n \in \mathbb{R}^p$ are i.i.d. sub-Gaussian variables with bounded variance and the noise vectors $\varepsilon_1,\hdots,\varepsilon_K$ are sub-Gaussian with variance proxy $\sigma^2$. Set $\lambda = C_\lambda \sigma\sqrt{\frac{\log(p-s)}{n}}$. In addition, assume that $|(\beta^*_0)_j|\le \gamma_1\,\forall j\in[p]$ and $|(\beta^*_k)_j|\le \gamma_2\,\forall k\in[K]\,\forall j\in[p]$ for some $\gamma_1\in(0,\infty)$ and $\gamma_2 \in (0,\frac{\lambda}{4})$, and the number of samples satisfy
\begin{align} \label{eq:randomdesign2}
    n\ge C_5 \max(1,\gamma_1^2)\, s \log\big(\max(p-s,K)\big),
\end{align}
and
\begin{align}
    \min_{j} |\big( \beta^*_0 \big )_j| \ge C_5^\prime \sigma \sqrt{\frac{\log(p-s)}{n}}.
    \label{eq:betamin_condition2}
\end{align}
Then, the pretraining estimator $\betapre$ exactly recovers the support of the common parameter $\beta_0^*$ with probability at least $1-C_6e^{-C_7n}$. Here, $C_\lambda, C_5,C_5^\prime,C_6$ are constants. 
\end{theorem}

\begin{remark}
\normalfont
\small
    We note that the above theorem imposes the condition $|\beta^*_k|\le \gamma_2\,\forall k\in[K]$ for some $\gamma_2 \le C_\lambda \sigma \sqrt{\frac{\log(p-s)}{n}}$ on the individual coefficients. This is a more stringent requirement compared to Theorem \ref{thm:recovery_model1} where $\gamma$ is unrestricted.
\end{remark}

\subsubsection{Error bounds for the two-stage procedure}

\label{sec:msebounds}
We now present an analysis of a simplified form of our pretraining strategy followed by the fitting of individual models.  Consider the common and individual support model 
\begin{align}
y_k=X_k \beta^*_0+X_k\beta^*_k \in \mathbb{R}^{n/K} + \varepsilon_k\quad \mbox{for }k\in[K].
\end{align}
Let us denote the full feature matrix $X=[X_1^T,...,X_k^T]^T \in\mathbb{R}^{n\times p}$. 
First, we fit our pretraining estimator 
\begin{align*}
    \hat \beta_0  \in \arg\min_{ \beta_0 \in \mathbb{R}^p} \sum_{k=1}^K \| X_k \beta_0 - y_ k\|_2^2 + \lambda \|\beta_0\|_1\,.
\end{align*}
%
Consequently, we fit the individual models using $\hat \beta_0$ as an offset term
\begin{align*}
    \hat \beta_k \in \arg\min_{ \beta_k \in \mathbb{R}^p}  \| X_k \hat \beta_0 + X_k\beta_k - y_k\|_2^2 + \lambda^\prime \|\beta_k\|_1\,,
\end{align*}
%
for $k=1,...,K$. Note that we omit the penalty factors and apply $\ell_1$ regularization without weighting. In addition, we assume that the tuning parameter $\alpha$ is fixed and known, and ignore it since it can be absorbed into the parameters. The following result presents the prediction error for this two-step procedure.

\begin{theorem}
\label{thm:prediction_error}
Suppose that $X_k\in\mathbb{R}^{n/K\times p}$ for $k\in[K]$ are fixed matrices and the noise vectors $\varepsilon_1,\hdots,\varepsilon_K$ are sub-Gaussian with variance proxy $\sigma^2$. Suppose that there exists constants $C,C^\prime$ such that the columns obey the average magnitude constraint $\frac{1}{n}\sum_{i=1}^{n}X_{ij}^2\le C $ for all $j\in[p]$, the average correlation constraint $\max_{j,j^\prime\in[p]}|\frac{1}{n/K} \sum_{i=1}^{n/K} (X_k)_{ij}(X_k)_{ij^\prime}|\le C^\prime$ for all $k\in[K]$. Let the regularization parameters satisfy $\lambda \ge \sqrt{n}{2R(2 C \sigma \sqrt{\log(p)}} +  C^\prime \frac{n}{K}R^\prime)$ and $\lambda^\prime \ge 4\sigma \sqrt{n/K}\sqrt{\log(pK)}$ where $R:=\|\beta_0^*\|_1$ and $R^\prime:=\sum_{k=1}^K \|\beta_k^*\|_1$.  Then, we have the following in-sample prediction error bound
\begin{align}
\frac{1}{n} \sum_{k=1}^{K} \|X_k ({\hat \beta_0-\beta_0^*} + \hat \beta_k - \beta_k^*) \|_2^2  
&\le
\frac{\lambda R + \lambda^\prime R^\prime}{n}+
\frac{\sigma(C R \sqrt{\log(p)}+ R^\prime \sqrt{\log(pK)/K})}{\sqrt{n}} 
+
\frac{C^\prime R^\prime}{K}\nonumber,
\end{align}
with probability at least $1-C_3/n$ for a certain constant $C_3$.

\end{theorem}
\begin{remark}
\small
   The prediction error is composed of three parts: the first part is proportional to the regularization parameters \(\lambda\) and \(\lambda^\prime\), which can be minimized by selecting the smallest permissible values, $\lambda = \sqrt{n}{2R(2 C \sigma \sqrt{\log(p)}} +  C^\prime \frac{n}{K}R^\prime)$ and $\lambda^\prime = 4\sigma \sqrt{n/K}\sqrt{\log(pK)}$. The second part decreases to zero as the total number of samples \(n\) grows to infinity, and the third part vanishes as the number of groups \(K\) increases. It is important to note that \(R^\prime = \sum_{k=1}^K \|\beta_k^*\|_1\), the sum of the \(\ell_1\) norms of the individual parameters \(\beta_k^*\), should grow more slowly than \(K\). For example, a scaling of \(\|\beta_k^*\|_1 = \mathcal{O}\left(\frac{1}{K}\right)\) for all \(k \in [K]\) satisfies this condition. Such a scaling assumption is unavoidable since we need to ensure that the pretraining stage estimates the common parameter $\beta_0$ in the presence of individual parameters which effectively act as a disturbance term.
\end{remark}
\begin{remark}
The above prediction error can be compared with the individual ordinary lasso estimators $\tilde \beta_k$ fitted to the data $X_k,y_k$ for each $k\in[K]$. A standard upper-bound for the average in-sample prediction error for this scheme under the same assumptions as in Theorem \ref{thm:prediction_error} is (e.g., see Theorem 11.2. in \cite{hastie2015statistical})
\begin{align}
    \frac{1}{n} \sum_{k=1}^K \| X_k  (\tilde \beta_k -\beta_0^* - \beta_k^*)\|_2^2 \lesssim  \frac{ (\sqrt{K} R+C^{\prime\prime}/\sqrt{K}) \sqrt{\log p}}{\sqrt{n}},
\end{align}
which holds with the probability at least $1-C_4/n$ for some constant $C_4$. We emphasize that the term $\sqrt{K} R$ is due to ignoring the common component $\beta_0$ across all groups, and leads to a factor of $\sqrt{K}$ larger prediction error compared to our bound provided in Theorem \ref{thm:prediction_error}.
\end{remark}
The proof of Theorem \ref{thm:prediction_error} is provided in the Appendix.

\section{Discussion}
\label{sec:discussion}
In this paper we have developed
a framework that enables the power of ML pretraining --- designed for neural nets --- to be applied in a simpler statistical setting (the lasso). We discuss many diverse applications of this paradigm, including
stratified models, multinomial responses, multi-response models, conditional average treatment estimation and  gradient boosting. Pretraining can be extended to incorporate unlabeled data alongside labeled data using sparse principal component analysis for feature selection. There are likely to be other interesting applications of these ideas. An open source R language package implementing this work is available at \href{https://github.com/erincr/ptLasso}{github.com/erincr/ptLasso} and will be on CRAN in the near future. 

\medskip
{\bf Acknowledgments.}
The authors would like to thank Emmanuel Candes, Daisy Ding, Trevor Hastie, Sarah McGough,  Vishnu Shankar, Lu Tian,  Ryan Tibshirani, and Stefan Wager for helpful discussions.
We thank Yu Liu and Yingcun Xia for implementing changes to their R package ODRF.
B.N.~was supported by National Center For Advancing Translational Sciences of the National Institutes of Health under Award Number UL1TR003142. M.A.R.~is in part supported by NHGRI under award R01HG010140, and by NIMH under award R01MH124244.
R.T.~was supported by the NIH (5R01EB001988-16) and the NSF (19DMS1208164). M.P. was supported in part by NSF under Grant DMS-2134248, in part by the NSF CAREER Award under Grant CCF-2236829, in part by the U.S. Army Research Office Early Career Award under Grant W911NF-21-1-0242, and in part by the Stanford Precourt Institute. E.C.~was supported by the Stanford Data Science Scholars Program and the Stanford Graduate Fellowship. J.S.~was supported by the NIGMS, Stanford University Discovery Innovation Award, and the Chan Zuckerberg Data Insights. S.G.~was supported by the Postdoctoral fellowship granted by the Western Norway Regional Health Authority.


\FloatBarrier

\bibliographystyle{plainnat}
\bibliography{reference}

\clearpage

\begin{appendices}

\section{Simulation study results}
\subsection{Input grouped data}
\label{app:simData}
Here, we consider the input grouped case, and we compare pretraining to the overall model and individual models in terms of (1) predictive performance on test data, (2) their F1 scores for feature selection and (3) F1 scores for feature selection among the common features only. Our simulations cover grouped data with a continuous response (Table ~\ref{tab:sim_gaussian}), grouped data with a binomial response (Table ~\ref{tab:sim_binomial}), and data with a multinomial response (Table~\ref{tab:sim_multinomial}).

\begin{table}
\small
    \centering
    \begin{tabular}{r|rrr|rrr|rr}
        \multicolumn{1}{c|}{SNR} & \multicolumn{3}{c|}{PSE relative to Bayes error}& \multicolumn{3}{c}{Feature F1} & \multicolumn{2}{|c}{Common feature F1}\\
          & Overall & Pretrain & Indiv. & 
          Overall & Pretrain & Indiv. & 
          Pretrain & Indiv.\\
         \toprule
         \multicolumn{9}{l}{Common support with same magnitude, individual features} \\
         \multicolumn{9}{l}{Features: $10$ common (coefficient values: 5), $10$ per group (coefficient values: 3), $120$ total} \\
        \midrule

        $13.6$ & $4.3 \pm 0.3$ & $\mathbf{1.4 \pm 0.1}$ & $1.4 \pm 0.1$ & $\mathbf{75 \pm 3}$ & $72 \pm 2$ & $70 \pm 1$ & $31 \pm 6$ & $\mathbf{39 \pm 5}$ \\

        $3.4$ & $1.9 \pm 0.1$ & $\mathbf{1.4 \pm 0.1}$ & $\mathbf{1.4 \pm 0.1}$ & $70 \pm 5$ & $\mathbf{73 \pm 3}$ & $70 \pm 1$ & $\mathbf{40 \pm 8}$ & $39 \pm 5$ \\

        $0.4$ & $\mathbf{1.1 \pm 0.0}$ & $1.2 \pm 0.0$ & $1.3 \pm 0.1$ & $23 \pm 4$ & $50 \pm 6$ & $\mathbf{64 \pm 5}$ & $\mathbf{86 \pm 10}$ & $73 \pm 14$ \\

        \midrule
        \multicolumn{9}{l}{As above, but now $p > n$} \\ 
        \multicolumn{9}{l}{Features: $10$ common (coefficient values: 5), $10$ per group (coefficient values: 2), $2040$ total} \\
        \midrule

        $11.6$ & $2.8 \pm 0.1$ & $\mathbf{1.9 \pm 0.2}$ & $2.7 \pm 0.2$ & $\mathbf{37 \pm 5}$ & $33 \pm 5$ & $21 \pm 2$ & $72 \pm 17$ & $\mathbf{91 \pm 6}$ \\

        $2.9$ & $\mathbf{1.5 \pm 0.1}$ & $1.5 \pm 0.1$ & $2.1 \pm 0.1$ & $\mathbf{33 \pm 4}$ & $31 \pm 5$ & $18 \pm 2$ & $81 \pm 15$ & $\mathbf{97 \pm 4}$ \\

        $0.3$ & $\mathbf{1.1 \pm 0.0}$ & $1.2 \pm 0.0$ & $1.3 \pm 0.0$ & $\mathbf{29 \pm 4}$ & $23 \pm 5$ & $13 \pm 4$ & $\mathbf{84 \pm 12}$ & $5 \pm 9$ \\
         
        \midrule
        \multicolumn{9}{l}{Common support with same magnitude, no individual features} \\
        \multicolumn{9}{l}{Features: $10$ common (coefficient values: 5), $0$ per group, $120$ total} \\ 
        \midrule

        $10.0$ & $\mathbf{1.0 \pm 0.0}$ & $1.1 \pm 0.0$ & $1.2 \pm 0.1$ & $50 \pm 11$ & $\mathbf{88 \pm 14}$ & $21 \pm 2$ & $\mathbf{92 \pm 7}$ & $76 \pm 9$ \\

        $2.5$ & $\mathbf{1.1 \pm 0.0}$ & $\mathbf{1.1 \pm 0.0}$ & $1.2 \pm 0.0$ & $52 \pm 11$ & $\mathbf{85 \pm 16}$ & $21 \pm 2$ & $\mathbf{91 \pm 9}$ & $74 \pm 9$ \\

        $0.3$ & $\mathbf{1.1 \pm 0.0}$ & $\mathbf{1.1 \pm 0.0}$ & $1.2 \pm 0.0$ & $51 \pm 11$ & $\mathbf{74 \pm 19}$ & $30 \pm 7$ & $\mathbf{92 \pm 8}$ & $76 \pm 15$ \\

         \midrule
        \multicolumn{9}{l}{Common support with different magnitudes, no individual features} \\ 
        \multicolumn{9}{l}{Features: $10$ common (coefficient values: 1 in group 1, 2 in group 2, etc.), $0$ per group, $120$ total} \\ 
        \midrule

         $4.4$ & $1.9 \pm 0.1$ & $\mathbf{1.1 \pm 0.0}$ & $1.2 \pm 0.1$ & $51 \pm 10$ & $\mathbf{62 \pm 20}$ & $22 \pm 2$ & $\mathbf{93 \pm 8}$ & $79 \pm 9$ \\

         $1.1$ & $1.3 \pm 0.0$ & $\mathbf{1.1 \pm 0.0}$ & $1.2 \pm 0.0$ & $52 \pm 10$ & $\mathbf{73 \pm 22}$ & $23 \pm 3$ & $\mathbf{93 \pm 9}$ & $86 \pm 10$ \\

         $0.1$ & $\mathbf{1.1 \pm 0.0}$ & $\mathbf{1.1 \pm 0.0}$ & $\mathbf{1.1 \pm 0.0}$ & $\mathbf{53 \pm 12}$ & $52 \pm 18$ & $38 \pm 11$ & $\mathbf{64 \pm 21}$ & $16 \pm 19$ \\
        
         \midrule
        \multicolumn{9}{l}{As above, but now with individual features} \\ 
        \multicolumn{9}{l}{Features: $10$ common (as above), $10$ per group  (coefficient values: 3), $120$ total} \\ 
        \midrule

        $10.8$ & $3.8 \pm 0.2$ & $\mathbf{1.4 \pm 0.1}$ & $\mathbf{1.4 \pm 0.1}$ & $64 \pm 5$ & $\mathbf{73 \pm 3}$ & $70 \pm 1$ & $\mathbf{49 \pm 9}$ & $40 \pm 5$ \\

        $4.8$ & $2.3 \pm 0.1$ & $\mathbf{1.3 \pm 0.1}$ & $1.4 \pm 0.1$ & $61 \pm 5$ & $\mathbf{73 \pm 3}$ & $70 \pm 2$ & $\mathbf{64 \pm 13}$ & $47 \pm 7$ \\

        $1.2$ & $1.4 \pm 0.1$ & $\mathbf{1.2 \pm 0.1}$ & $1.3 \pm 0.1$ & $55 \pm 5$ & $51 \pm 10$ & $\mathbf{65 \pm 3}$ & $\mathbf{86 \pm 11}$ & $76 \pm 12$ \\
        
         \midrule
        \multicolumn{9}{l}{Individual features only} \\ 
        \multicolumn{9}{l}{Features: $0$ common, $10$ per group  (coefficient values: 5), $120$ total} \\ 
        \midrule

         $10.1$ & $9.8 \pm 0.6$ & $\mathbf{1.2 \pm 0.0}$ & $\mathbf{1.2 \pm 0.0}$ & $65 \pm 3$ & $\mathbf{67 \pm 3}$ & $\mathbf{67 \pm 4}$ & ---  & --- \\

        $2.5$ & $3.3 \pm 0.2$ & $\mathbf{1.2 \pm 0.1}$ & $\mathbf{1.2 \pm 0.1}$ & $63 \pm 6$ & $\mathbf{68 \pm 3}$ & $\mathbf{68 \pm 2}$ & --- & --- \\

        $0.6$ & $1.6 \pm 0.1$ & $\mathbf{1.3 \pm 0.1}$ & $\mathbf{1.3 \pm 0.1}$ & $52 \pm 5$ & $\mathbf{69 \pm 2}$ & $68 \pm 3$ & --- & --- \\
        
         \bottomrule
    \end{tabular}
    \caption{ \em Gaussian response, 5 input groups. Each input group has 100 observations; in total there are $n = 500$ observations. Each row represents $100$ simulations, and each result shows the mean and one standard deviation.}
    \label{tab:sim_gaussian}
\end{table}

\begin{table}
\small
    \centering
    \begin{tabular}{rrrr|rrr|rr}
    \multicolumn{4}{c|}{Test AUC (x 100)} & \multicolumn{3}{c}{Feature F1} & \multicolumn{2}{c}{Common feature F1} \\
          Bayes & Overall & Pretrain & Indiv. & 
          Overall & Pretrain & Indiv. & 
          Pretrain & Indiv. \\
         \toprule
         \multicolumn{9}{l}{Common support with same magnitude, individual features} \\
         \multicolumn{9}{l}{Features: 5 common (-.5, .5, .3, -.9, .1), 5 per group ( -.45, .45,  .27, -.81,  .09), 40 total} \\
         \midrule
        $82 \pm 2$ & $71 \pm 3$ & $\mathbf{72 \pm 4}$ & $70 \pm 4$ & $62 \pm 6$ & $65 \pm 8$ & $\mathbf{66 \pm 6}$ & $\mathbf{44 \pm 15}$ & $39 \pm 13$ \\

         $0.7 \pm 3$ & $\mathbf{60 \pm 4}$ & $59 \pm 4$ & $58 \pm 3$ & $57 \pm 8$ & $60 \pm 11$ & $\mathbf{63 \pm 8}$ & $\mathbf{35 \pm 12}$ & $28 \pm 12$ \\
        
        $61 \pm 3$ & $\mathbf{54 \pm 3}$ & $53 \pm 3$ & $53 \pm 3$ & $57 \pm 9$ & $\mathbf{60 \pm 13}$ & $\mathbf{60 \pm 13}$ & $\mathbf{24 \pm 13}$ & $18 \pm 11$ \\
         
         \midrule
         \multicolumn{9}{l}{As above, but now $p > n$} \\
         \multicolumn{9}{l}{Features: 5 common (-.5, .5, .3, -.9, .1), 5 per group ( -.45, .45,  .27, -.81,  .09), 320 total} \\
         \midrule

         $82 \pm 2$ & $\mathbf{68 \pm 4}$ & $67 \pm 4$ & $63 \pm 4$ & $\mathbf{42 \pm 9}$ & $24 \pm 7$ & $23 \pm 6$ & $\mathbf{40 \pm 16}$ & $32 \pm 13$ \\

          $70 \pm 3$ & $\mathbf{56 \pm 4}$ & $55 \pm 4$ & $53 \pm 4$ & $\mathbf{29 \pm 11}$ & $16 \pm 6$ & $15 \pm 5$ & $\mathbf{24 \pm 14}$ & $14 \pm 13$ \\

          $61 \pm 3$ & $\mathbf{51 \pm 3}$ & $\mathbf{51 \pm 3}$ & $\mathbf{51 \pm 4}$ & $\mathbf{21 \pm 9}$ & $10 \pm 6$ & $12 \pm 4$ & $\mathbf{8 \pm 8}$ & $3 \pm 6$ \\
         
         \midrule
        \multicolumn{9}{l}{Common support with same magnitude, no individual features} \\ 
        \multicolumn{9}{l}{Features: 5 common (-.5, .5, .3, -.9, .1), 0 per group, 40 total} \\
        \midrule

        $77 \pm 2$ & $\mathbf{73 \pm 3}$ & $70 \pm 4$ & $66 \pm 4$ & $\mathbf{56 \pm 7}$ & $47 \pm 16$ & $\mathbf{56 \pm 14}$ & $\mathbf{63 \pm 16}$ & $50 \pm 17$ \\

        $65 \pm 3$ & $\mathbf{60 \pm 4}$ & $57 \pm 4$ & $55 \pm 4$ & $54 \pm 9$ & $54 \pm 16$ & $\mathbf{62 \pm 11}$ & $\mathbf{39 \pm 15}$ & $30 \pm 13$ \\

        $58 \pm 4$ & $\mathbf{53 \pm 4}$ & $51 \pm 4$ & $51 \pm 3$ & $55 \pm 9$ & $59 \pm 14$ & $\mathbf{63 \pm 9}$ & $\mathbf{27 \pm 11}$ & $23 \pm 10$ \\
        
        \midrule
        \multicolumn{9}{l}{Common support with different magnitudes, no individual features} \\ 
        \multicolumn{9}{l}{Features: 5 common, 0 per group, 40 total} \\
        \multicolumn{9}{l}{\phantom{Features: }Group 1 coefficient values: (-.5, .5, .3, -.9, .1)} \\
        \multicolumn{9}{l}{\phantom{Features: }Group 2 coefficient values: (-.3, .9, .1, -.1, .2)} \\
        \multicolumn{9}{l}{\phantom{Features: }Group 3 coefficient values: (.1, .2, -.1, .2, .3)} \\
        \midrule

        $71 \pm 3$ & $61 \pm 4$ & $\mathbf{62 \pm 4}$ & $61 \pm 4$ & $53 \pm 9$ & $\mathbf{56 \pm 14}$ & $\mathbf{56 \pm 15}$ & $\mathbf{44 \pm 16}$ & $36 \pm 17$ \\

        $62 \pm 3$ & $\mathbf{54 \pm 4}$ & $\mathbf{54 \pm 4}$ & $53 \pm 4$ & $56 \pm 9$ & $60 \pm 15$ & $\mathbf{62 \pm 11}$ & $\mathbf{29 \pm 11}$ & $27 \pm 10$ \\

        $56 \pm 4$ & $51 \pm 3$ & $\mathbf{52 \pm 3}$ & $51 \pm 3$ & $56 \pm 10$ & $61 \pm 13$ & $\mathbf{64 \pm 5}$ & $\mathbf{22 \pm 11}$ & $\mathbf{22 \pm 10}$ \\
        
        \midrule
        \multicolumn{9}{l}{As above, but now with individual features} \\  
        \multicolumn{9}{l}{Features: 5 common, 5 per group, 40 total} \\
        \multicolumn{9}{l}{\phantom{Features: }Common support coefficients are as above.} \\
        \multicolumn{9}{l}{\phantom{Features: }Individual support coefficient values are 0.9 $ \times $ common support coefficient values.} \\
        \midrule
        $76 \pm 3$ & $61 \pm 4$ & $\mathbf{65 \pm 4}$ & $64 \pm 4$ & $58 \pm 9$ & $63 \pm 9$ & $\mathbf{64 \pm 8}$ & $\mathbf{37 \pm 14}$ & $30 \pm 15$ \\

        $70 \pm 3$ & $56 \pm 4$ & $\mathbf{58 \pm 4}$ & $\mathbf{58 \pm 4}$ & $56 \pm 10$ & $61 \pm 8$ & $\mathbf{63 \pm 7}$ & $\mathbf{28 \pm 13}$ & $26 \pm 13$ \\

        $64 \pm 3$ & $\mathbf{54 \pm 3}$ & $\mathbf{54 \pm 4}$ & $\mathbf{54 \pm 3}$ & $55 \pm 9$ & $58 \pm 14$ & $\mathbf{61 \pm 12}$ & $\mathbf{24 \pm 13}$ & $21 \pm 12$ \\
        
        \midrule
        \multicolumn{9}{l}{Individual features only} \\ 
        \multicolumn{9}{l}{Group 1 coefficient values: (-.5, .5, .3, -.9, .1)} \\
        \multicolumn{9}{l}{Group 2 coefficient values: (-.3, .9, .1, -.1, .2)} \\
        \multicolumn{9}{l}{Group 3 coefficient values: (.1, .2, -.1, .2, .3)} \\
        \midrule

        $72 \pm 3$ & $56 \pm 3$ & $\mathbf{62 \pm 4}$ & $\mathbf{62 \pm 4}$ & $\mathbf{57 \pm 9}$ & $59 \pm 10$ & $55 \pm 14$ & --- & --- \\

        $66 \pm 3$ & $53 \pm 4$ & $\mathbf{56 \pm 3}$ & $\mathbf{56 \pm 3}$ & $58 \pm 9$ & $\mathbf{60 \pm 12}$ & $59 \pm 13$ & --- & --- \\

         $60 \pm 3$ & $52 \pm 3$ & $\mathbf{53 \pm 3}$ & $\mathbf{53 \pm 3}$ & $57 \pm 8$ & $\mathbf{63 \pm 10}$ & $59 \pm 14$ & --- & --- \\
        
         \bottomrule
    \end{tabular}
    \caption{\em Binomial response, 3 input groups. Each input group has 100 observations; in total there are $n = 300$ training observations. Each row represents $100$ simulations, and each result shows the mean and one standard deviation.}
    \label{tab:sim_binomial}
\end{table}

\begin{table}[ht]
\small
    \centering
    \begin{tabular}{rrrr|rrr|rr}
    \multicolumn{4}{c|}{Test misclassification rate (x 100)} & \multicolumn{3}{c}{Feature F1} & \multicolumn{2}{c}{Common feature F1} \\
          Bayes rule & Overall & Pretrain & Indiv. & 
          Overall & Pretrain & Indiv. & 
          Pretrain & Indiv. \\
         \toprule
         \multicolumn{9}{l}{Common support, individual features} \\ 
         \multicolumn{9}{l}{Features: 3 common (simulation 1: group 1 - group 5 values: -.5, -0.25, 0, 0.25, 0.5)} \\ 
         \multicolumn{9}{l}{\phantom{Features: 3 common }(simulation 2: group 1 - group 5 values: -1, -.5, 0, 0.5, 1)} \\
         \multicolumn{9}{l}{\phantom{Features: }10 per group (same values as above)} \\ 
         \multicolumn{9}{l}{\phantom{Features: }159 total} \\ 
         \midrule
         $49 \pm 1$ & $59 \pm 2$ & $\mathbf{58 \pm 2}$ & $60 \pm 2$ & $55 \pm 8$ & $\mathbf{57 \pm 5}$ & $55 \pm 5$ & $26 \pm 19$ & $\mathbf{28 \pm 31}$ \\

         $22 \pm 1$ & $33 \pm 2$ & $\mathbf{32 \pm 2}$ & $\mathbf{32 \pm 3}$ & $\mathbf{71 \pm 6}$ & $65 \pm 6$ & $65 \pm 6$ & $20 \pm 15$ & $\mathbf{55 \pm 20}$ \\

         \midrule
         \multicolumn{9}{l}{As above, but now $p > n$} \\
         \multicolumn{9}{l}{Features: 3 common, 10 per group, 640 total} \\ 
         \multicolumn{9}{l}{\phantom{Features: }Values as above, with extra noise features.} \\ 
         \midrule
         $56 \pm 1$ & $62 \pm 2$ & $\mathbf{61 \pm 2}$ & $62 \pm 3$ & $34 \pm 7$ & $\mathbf{36 \pm 6}$ & $\mathbf{36 \pm 5}$ & $\mathbf{26 \pm 20}$ & $21 \pm 30$ \\

          $28 \pm 1$ & $35 \pm 3$ & $\mathbf{34 \pm 2}$ & $36 \pm 4$ & $\mathbf{56 \pm 11}$ & $48 \pm 8$ & $49 \pm 7$ & $24 \pm 22$ & $\mathbf{73 \pm 21}$ \\
         \midrule
        \multicolumn{9}{l}{Common support, no individual features} \\ 
        \multicolumn{9}{l}{Features: 3 common, 0 per group, 159 total} \\ 
        \multicolumn{9}{l}{\phantom{Features: }Common feature values as above.} \\ 
        \midrule
         $70 \pm 1$ & $71 \pm 3$ & $\mathbf{70 \pm 2}$ & $75 \pm 4$ & $\mathbf{32 \pm 13}$ & $28 \pm 10$ & $21 \pm 11$ & $\mathbf{36 \pm 22}$ & $7 \pm 21$ \\

         $58 \pm 1$ & $\mathbf{57 \pm 2}$ & $58 \pm 1$ & $60 \pm 2$ & $24 \pm 10$ & $29 \pm 10$ & $\mathbf{31 \pm 10}$ & $\mathbf{49 \pm 24}$ & $36 \pm 45$ \\

        \midrule
        \multicolumn{9}{l}{Individual features only} \\
        \multicolumn{9}{l}{Features: 0 common, 10 per group, 159 total} \\
        \multicolumn{9}{l}{\phantom{Features: }Individual feature values as above.} \\
        \midrule
        $54 \pm 1$ & $64 \pm 2$ & $\mathbf{63 \pm 2}$ & $64 \pm 3$ & $\mathbf{54 \pm 7}$ & $\mathbf{54 \pm 4}$ & $50 \pm 7$ & --- & --- \\

        $27 \pm 1$ & $38 \pm 2$ & $\mathbf{36 \pm 2}$ & $37 \pm 3$ & $\mathbf{69 \pm 6}$ & $63 \pm 6$ & $64 \pm 6$ & --- & --- \\
         \bottomrule
    \end{tabular}
    \caption{\em Multinomial response, 5 input groups. Each input group has 50 observations; in total there are $n = 250$ observations. Each row represents $100$ simulations, and each result shows the mean and one standard deviation.}
    \label{tab:sim_multinomial}
\end{table}
\FloatBarrier

\subsection{External dataset}
\label{app:simDataExternal}

\begin{figure}
\centering
\includegraphics[width=.9 \linewidth]{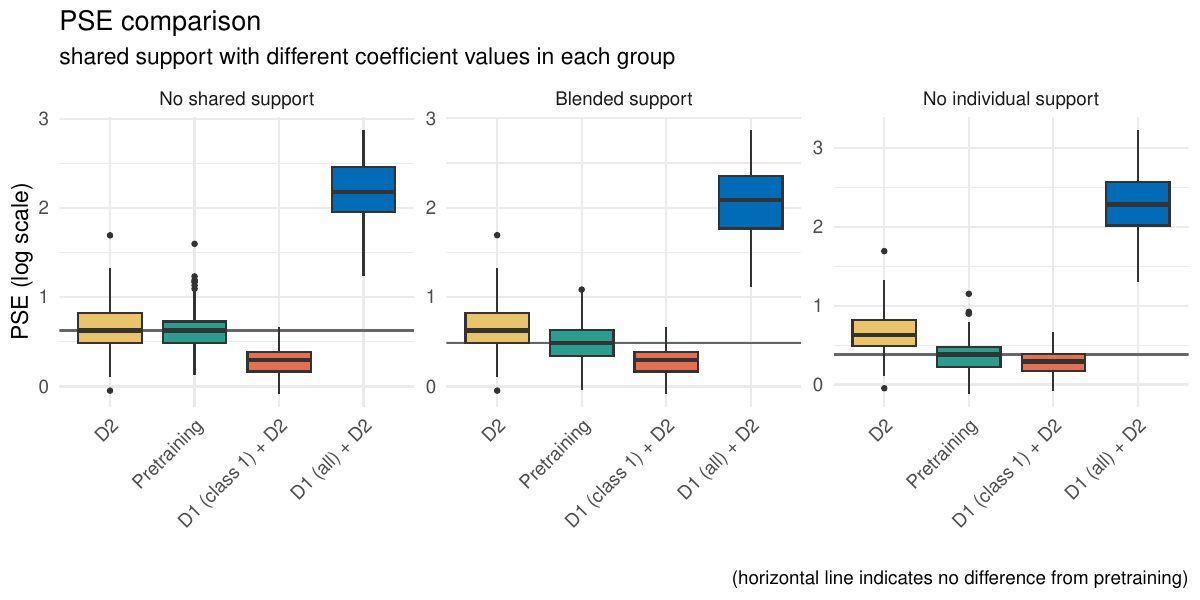}
\caption{\em Comparison of approaches for modeling with a large (usually inaccessible) dataset $D_1$ and a smaller dataset $D_2$. When $D_1$ is informative, pretraining using only $D_2$ and the coefficients from a model trained with $D_1$ outperforms using $D_2$ alone. When $D_1$ is not informative, pretraining with $D_1$ does not hurt.}
\label{fig:external2}
\end{figure}

We repeat the experiment in Section~\ref{sec:external}, now studying the setting where the $10$ groups in $D_1$ share support, but the coefficients on that support have different values (same sign). As before, the group specific coefficients $\beta_g$ were a mixture of shared and group-specific coefficients, controlled by a parameter $\rho$: 
\[\beta_g = \rho \times \beta_{g,\text{ shared}} + (1 - \rho) \times \beta_{g,  \text{ indiv}},\]  
where $\beta_{g,\text{ shared}}$ is now a vector with the same support (non-zero entries), but different values for each group, and $\beta_{g,  \text{ indiv}}$ is group specific. Also as before, the supports for $\beta_{g, \text{ indiv}}$ are non-overlapping across groups. Performance for $\rho = 0, 0.5, 1$ is shown in Figure~\ref{fig:external2}. Using all of $D_1$ hurts performance enough that the plot becomes difficult to read; we show the plot again with this case omitted. Pretraining always matched or outperformed using $D_2$ alone; access to class $1$ from $D_1$ performed better than or roughly the same as pretraining.

\section{Mathematical Proofs}
\subsection{Proof of Lemma \ref{lem:irrep_cond}}
We analyze the conditions for optimality of the pretraining estimator given in \eqref{eq:pretraining_lasso}. We apply the scaling $X\leftarrow \frac{1}{\sqrt{n}} X$ and $y\leftarrow \frac{1}{\sqrt{n}} y$ to absorb the $\frac{1}{n}$ factor and simplify our notation. Suppose that the support of the optimal solution $\beta$ is $S$, which is assumed to contain the support of $\beta^*_k~\forall k$.  The optimality conditions that ensure $\beta$ is the unique solution with support $S$ are as follows

\begin{align}
    X_S^T(X_S{\beta}_S-y) + \lambda \signb(\beta_S) &= 0\\
    \|X_{S^c}^T(X_S{\beta}_S-y)\|_{\infty}&< \lambda,\label{eq:irrep_second_proof}
\end{align}
where $\beta = \betapre$ is the optimal solution. When the matrix $X_S\in\mathbb{R}^{n\times s}$ is full column-rank,  the matrix $X_S^T X_S$ is invertible and we can solve for $\beta_S$ as follows

\begin{align}
\label{eq:betasoln}
    \beta_S = (X_S^TX_S)^{-1} (X_S^Ty - \lambda \signb(\beta_S)).
\end{align}

Plugging in the observation model $y= \sum_{k=1}^K D_k Xw^*_k + \varepsilon$, we obtain

\begin{align}
\label{eqn:beta_expression_y}
    \beta_S = (X_S^TX_S)^{-1} (X_S^T\sum_{k=1}^K D_k Xw^*_k) + X_S^\dagger \varepsilon - \lambda (X_S^TX_S)^{-1}\signb(\beta_S).
\end{align}

Plugging in the above expression into the condition \eqref{eq:irrep_second_proof}, and dividing both sides by $\lambda$, we obtain 

\begin{align}
    \|X_{S^c}^T (\lambda^{-1}P_S^\perp  \sum_{k=1}^K D_k Xw^*_k + \lambda^{-1} P_S^\perp \varepsilon +  X_S^\dagger \signb(\beta_S) \|_{\infty}< 1.
\end{align}
Using triangle inequality, we upper-bound the left-hand-side to arrive the sufficient condition
\begin{align}
    \lambda^{-1} \|X_{S^c}^T P_S^\perp  \sum_{k=1}^K D_k Xw^*_k\|_{\infty} +  \lambda^{-1} \|X_{S^c}^T P_S^\perp  \varepsilon\|_{\infty} + \|X_{S^c}^T X_S^\dagger \signb(\beta_S) \|_{\infty}
    < 1.
\end{align}

Therefore by imposing the conditions
\begin{align}
    \|X_{S^c}^T X_S^\dagger \signb(\beta_S) \|_{\infty} &< \frac{1}{2}\\
    \|X_{S^c}^T P_S^\perp  \sum_{k=1}^K D_k Xw^*_k\|_{\infty}&\le \frac{ \lambda}{4}\\
    \|X_{S^c}^T P_S^\perp  \varepsilon\|_{\infty} &\le \frac{  \lambda}{4},
\end{align}
we observe that the optimality conditions for $\beta$ with the support $S$ are satisfied.

\subsection{Proof of Theorem \ref{thm:recovery_model1}}
\noindent\textbf{First condition}\\
We consider the first condition of pretraining irrepresentability given by
\begin{align}
    \|X_{S^c}^T X_S^\dagger \signb(\beta_S) \|_{\infty} = \max_{j \in S^c} |x_j^T X_S^\dagger \signb(\beta_S) |.
\end{align}
Note that $x_j^T$ and  $X_S^\dagger \signb(\beta_S)$ are independent for $j\in S^c$. Therefore, $X_S^\dagger \signb(\beta_S)$ is sub-Gaussian with variance proportional to $\frac{1}{n}\|X_S^\dagger \signb(\beta_S)\|_2^2$.

When $n\ge C s$ for some constant $C$, the matrix $X_S^TX_S$ is a near-isometry in spectral norm, i.e.,
\begin{align}
    \|X_S^TX_S-I\|_2\le \delta,
\end{align}
with probability at least $1-C_1e^{-C_2n}$ where $C_1,C_2$ are constants.
Therefore for $\delta<1$, we have $\|X_S^\dagger\|_2\le (1-\delta)^{-1}$ and $\|X_S^\dagger \signb(\beta_S)\|_2^2\lesssim \|\signb(\beta_S)\|_2^2=s$.

Applying union bound, we obtain
\begin{align}
    \mathbb{P} \left[ \max_{j \in S^c} |x_j^T X_S^\dagger \signb(\beta_S) | \le \delta  \right] &\le (p-s) 
     \mathbb{P} \left[  |x_1^T X_S^\dagger \signb(\beta_S) |\le \delta  \right]\\
     & \le (p-s) e^{-\delta^2C^\prime n/s }\\
     & = e^{-C^\prime \delta^2 n/s +\log(p-s)},
\end{align}
for some constant $C^\prime$. 

Consequently, for $n\gtrsim \delta^{-2}s \log(p-s)$ we have $\max_{j \in S^c} |x_j^T X_S^\dagger \signb(\beta_S) |\le \delta$ with probability at least $1-C_3e^{-C_4\delta^2n/s}$ where $C_3,C_4$ are constants.
~\\
~\\
\noindent\textbf{Second condition}\\
We proceed bounding the second irrepresentability condition involving the matrix $X_{S^c}^T P_S^\perp  \sum_{k=1}^K D_k Xw^*_k$ using the same strategy used above.
Note the critical fact that the shared support model implies the matrices $X_{S^c}$ and 
\begin{align*}
    P_S^\perp  \sum_{k=1}^K D_k Xw^*_k = P_S^\perp  \sum_{k=1}^K D_k X_S(w^*_k)_S,
\end{align*}
are independent since the latter matrix only depends on the features $X_S$. 

Note that 
\begin{align}
    \|P_S^\perp  \sum_{k=1}^K D_k X_S(w^*_k)_S\|_2 &\le \| \sum_{k=1}^K D_k X_S(w^*_k)_S\|_2\\
    &= \left(\sum_{k=1}^K \|  D_k X_S(w^*_k)_S\|_2^2\right)^{1/2}.
\end{align}

Recalling the scaling of the $X$ by $\frac{1}{n}$, we note that $D_kX_S$ is an $n\times s$ formed by the concatenation of an an $\frac{n}{K}\times s$ matrix of i.i.d. sub-Gaussian variables with variance $\mathcal{O}(\frac{1}{n})$ with an $(n-k)\times s$ matrix of zeros. From standard results on the singular values of sub-Gaussian matrices \cite{vershynin2018high}, we have $\|D_kX_S\|_2\lesssim \frac{\sqrt{n/K}+\sqrt{s}}{\sqrt{n}} =  \sqrt{\frac{1}{K}} + \sqrt{\frac{s}{n}} $ with probability at least $1-C_5e^{-C_6n}$. Using the fact that $w^*_k$ has entries bounded in $[-\gamma, +\gamma]$, we obtain the upper-bound
\begin{align}
    \|P_S^\perp  \sum_{k=1}^K D_k X_S(w^*_k)_S\|_2 &\le \left(\sum_{k=1}^K \|  D_k X_S(w^*_k)_S\|_2^2 \|(w^*_k)_S\|_2^2\right)^{1/2}\\
    &\lesssim \left(\sum_{k=1}^K (\frac{1}{K}+\frac{s}{n}) s \gamma^2 \right)^{1/2}\\
    &= \left( (1+\frac{Ks}{n}) s \gamma^2 \right)^{1/2}\\
    &\le \sqrt{s}\gamma + \frac{\sqrt{K}}{\sqrt{n}}s\gamma\\
    &\le 2\sqrt{s}\gamma,
\end{align}
where we used the fact that $n/K\ge s$, i.e., each subgroup has at least $s$ samples, in the final inequality.
Repeating the same argument involving sub-Gaussian variables and the union bound used for the first condition above, we obtain that for $n\gtrsim \delta^{-2}\gamma^2 s \log(p-s)$ we have $X_{S^c}^T P_S^\perp  \sum_{k=1}^K D_k Xw^*\le \delta$ with probability at least $1-C_7e^{-C_8n\delta^2/(s\gamma^2)}$ where $C_7,C_8$ are constants.

~\\
~\\
\noindent\textbf{Third condition}\\
Using standard results on Gaussian vectors, and repeating the union bound argument used in analyzing the first condition, we obtain that $ \|X_{S^c}^T P_S^\perp  \varepsilon\|_{\infty}\le \delta \lambda $ when we set $\lambda = C_\lambda \sigma \sqrt{\frac{\log{p-s}}{n}}$ and $n\gtrsim \delta^{-2} \sigma^2 \log(p-s)$ with probability at least $1-C_9e^{-C_{10}n\delta^2/\sigma^2}$ where $C_\lambda, C_9,C_{10}$ are constants.

Applying union bound to bound the probability that all of the three conditions hold simultaneously, we complete the proof of the theorem.
\subsection{Proof of Lemma \ref{lemma:isometry}}

We apply well-known concentration bounds for the extreme singular values of i.i.d. Gaussian matrices (see e.g. \cite{vershynin2018high}). These bounds $\|X_S^TX_S-K I\|\le c_1\delta $ and $\|X_S^TD_kX_S- I\|\le c_2 \delta$ for each fixed $k\in[K]$ with high probability when $n\succeq \delta^{-2} s$. Applying union bound over $k\in[K]$, we obtain the claimed result. 

\subsection{Proof of Lemma \ref{lemma:approx_avg}}
We consider the expression for $\beta_S$ given in \eqref{eqn:beta_expression_y} in the proof of Lemma \ref{lem:irrep_cond}. Applying triangle inequality to control the terms on the right-hand-side, we obtain the claimed result.
\subsection{Proof of Theorem \ref{thm:exact_recovery_model1}}
Note that we only need to control the signs of $\beta_S$ given in \eqref{eq:betasoln}, in addition to the guarantees of Theorem \ref{thm:recovery_model1}. Our strategy is to bound the $\ell_\infty$ norm of $\beta_S-\frac{1}{K} \sum_{k=1}^K \beta^*_k$ via its $\ell_2$ norm and establishing entrywise control on $\beta_S$ by the assumption on the minimum value of the average $\frac{1}{K} \sum_{k=1}^K \beta^*_k$. We combine Lemma \ref{lemma:isometry} and Lemma \ref{lemma:approx_avg} with the expression $\eqref{eq:betasoln}$ to obtain
\begin{align}
    \|\beta_S - \frac{1}{K} \sum_{k=1}^K \beta^*_k\|_\infty \le \lambda \|(X_S^TX_S)^{-1} \mathbf{sign}(\beta_S^*)\|_2 + \delta \sum_{k=1}^K \|\beta_k^*\|_2 +  \| X_S^\dagger \varepsilon \|,
\end{align}
with high probability.
Noting that $\|(X_S^TX_S)^{-1}\|_2\lesssim K(1+\delta)$ with high probability, and $\sum_{k=1}^K\|\beta_k^*\|_2\le K\gamma$ by our assumption on the magnitude of $\beta_k^*$, we obtain the claimed result.

\subsection{Proof of Theorem \ref{thm:recovery_model2}}
The main difference of this result compared to the proof of Theorem \ref{thm:recovery_model1} is in the analysis of the quantity $\|X_{S^c}^T P_S^\perp  \sum_{k=1}^K D_k Xw^*_k\|_{\infty}$. Unfortunately, $X_{S^c}$ and $P_S^\perp  \sum_{k=1}^K D_k Xw^*_k$ are no longer independent. We proceed as follows

\begin{align}
    \|X_{S^c}^T P_S^\perp  \sum_{k=1}^K D_k Xw^*_k\|_{\infty} &= \max_{j\in S^c}  |x_j^T\sum_{k=1}^K D_k Xw^*_k|\\
    & = |x_j^T\sum_{r\neq j}\sum_{k=1}^K D_k X(w^*_k)_r + x_j^TD_kX(w_k^*)_j|\\
    & \le  |x_j^T\sum_{r\neq j}\sum_{k=1}^K D_k X(w^*_k)_r| + |x_j^Tx_j(w_k^*)_j|
\end{align}
we bound the last term via $|x_j^Tx_j(w_k^*)_j|\le \|x_j\|_2^2 \gamma_2^2$ and impose $\gamma_2\in(0,\frac{\lambda}{4})$. Note that $\|x_j\|_2^2\lesssim 1$ with high probability due to the rescaling by $\frac{1}{n}$. The rest of the proof is identical to the proof of Theorem \ref{thm:recovery_model1}.

\subsection{Proof of Theorem \ref{thm:prediction_error}}
    We first derive an error bound for the pretraining stage using the basic inequality
    \begin{align}
        \sum_{k=1}^K \| X_k \hat \beta_0 - y_ k\|_2^2 + \lambda \|\hat \beta_0\|_1 \le \sum_{k=1}^K \| X_k \beta^*_0 - y_ k\|_2^2+ \lambda \|\beta_0^*\|_1 ,
    \end{align}
    which follows from the optimality of $\hat \beta_0$ in the pretraining lasso objective.
    
    Plugging in the model for $y$, we obtain
        \begin{align}
        \sum_{k=1}^K \| X_k (\hat \beta_0-\beta_0^*) - X_k\beta_k^*-\varepsilon_k\|_2^2 \le \sum_{k=1}^K \| X_k \beta^*_0 + \varepsilon_k\|_2^2 + \lambda (\|\beta_0^*\|_1-\|\hat\beta_0\|_1).
    \end{align}
    Expanding the square and cancelling common terms we get
    \begin{align}
        \sum_k \|X_k \Delta_0\|_2^2 \le 2 \sum_k (X_k\beta_k^*+\varepsilon_k)^T X_k \Delta_0+ \lambda (\|\beta_0^*\|_1-\|\hat\beta_0\|_1),
    \end{align}
    where we defined $\Delta_0:=\hat \beta_0-\beta_0^*.$ 
    We apply Cauchy–Schwarz inequality and triangle inequality to obtain
    \begin{align}
        \sum_{k=1}^K \|X_k \Delta_0\|_2^2 &\le 2 \|\Delta_0\|_1  \Big(\big\|\sum_k X_k^T\varepsilon_k\big\|_\infty + \sum_{k}\| X_k^TX_k \beta_k^*\|_{\infty}\Big)+\lambda (\|\beta_0^*\|_1-\|\hat\beta_0\|_1)\\
        &\le 2 \|\Delta_0\|_1  \Big(\big\|\sum_k X_k^T\varepsilon_k\big\|_\infty + \sum_{k}\| X_k^TX_k \beta_k^*\|_{\infty}\Big)+\lambda (\|\beta_0^*\|_1-\|\hat\beta_0\|_1)\\
        &\le 2 \|\Delta_0\|_1  \Big(\big\|\sum_k X_k^T\varepsilon_k\big\|_\infty + \sum_{k}\| X_k^TX_k\|_{\infty} \|\beta_k^*\|_{1}\Big)+\lambda (\|\beta_0^*\|_1-\|\hat\beta_0\|_1)\\
        &\le 2 \|\Delta_0\|_1  \Big(\big\|\sum_k X_k^T\varepsilon_k\big\|_\infty + \max_{k\in[K]}\max_{i,j\in[p]} |(X_k^TX_k)_{ij}| \sum_{k} \|\beta_k^*\|_{1}\Big)+\lambda (\|\beta_0^*\|_1-\|\hat\beta_0\|_1)\\
        &\le 2 \|\Delta_0\|_1  \Big(\big\|\sum_k X_k^T\varepsilon_k\big\|_\infty + \frac{C^\prime n}{K}\sum_{k} \|\beta_k^*\|_{1}\Big)+\lambda (\|\beta_0^*\|_1-\|\hat\beta_0\|_1)\\
        &\le 2 (\|\beta_0^*\|_1+\|\hat\beta_0\|_1) \Big(\big\|\sum_k X_k^T\varepsilon_k\big\|_\infty + \frac{C^\prime n}{K}\sum_{k} \|\beta_k^*\|_{1}\Big)+\lambda (\|\beta_0^*\|_1-\|\hat\beta_0\|_1).
    \end{align}
    Next, we pick $\lambda \ge 2\big\|\sum_k X_k^T\varepsilon_k\big\|_\infty + 2\frac{C^\prime n}{K}\sum_{k} \|\beta_k^*\|_{1}$ to obtain the the upper bound on the normalized prediction error

    $$ \frac{1}{n} \sum_{k=1}^K \|X_k \Delta_0\|_2^2 \le 2 \|\beta_0^*\|_1 \Big( \frac{1}{n}\big\|\sum_k X_k^T\varepsilon_k\big\|_\infty + \frac{C^\prime}{K}\sum_{k} \|\beta_k^*\|_{1}\Big)+\frac{\lambda}{n} \|\beta_0^*\|_1.$$
      
    We apply standard concentration results for the maximum of independent sub-Gaussian variables \cite{vershynin2018high} to control the term $\|\sum_k X_k^T\varepsilon_k\|_\infty$. After simplification we obtain
    \begin{align}
    \label{eqn:error_pretrain}
         \frac{1}{n}\sum_k \|X_k \Delta_0\|_2^2 &\le \frac{4\|\beta_0^*\|_1 C \sigma \sqrt{\log(p)}}{\sqrt{n}} + \frac{2 \|\beta_0^*\|_1 C^\prime \sum_k \|\beta_k^*\|_1}{K}+\frac{\lambda}{n} \|\beta_0^*\|_1\\
         &\le \|\beta_0^*\|_1\left(\frac{4 C \sigma \sqrt{\log(p)}}{\sqrt{n}} + \frac{2 C^\prime \sum_k \|\beta_k^*\|_1}{K}+\frac{\lambda}{n} \right)\,,
    \end{align}
    with probability at least $1-C_3/n$ when we set $\lambda \ge \sqrt{n}{2\|\beta_0^*\|_1(2 C \sigma \sqrt{\log(p)}} +  C^\prime \frac{n}{K}\sum_k \|\beta_k^*\|_1)$.
    The above inequality shows that the prediction error of the pretraining stage, $X\hat \beta_0 - X\beta_0^*$ is controlled with high probability.
    
    Next, we analyze the second stage using the same basic inequality argument used above. We have
    \begin{align}
    \| X_k \hat \beta_0 + X_k\hat \beta_k - y_k\|_2^2 \le \| X_k \hat \beta_0 + X_k\beta_k^* - y_k\|_2^2 + \lambda^\prime(\|\beta_k^*\|_1-\|\hat\beta_k\|_1).
    \end{align}
    Defining $\Delta_k:=\hat \beta_k - \beta_k^*$ for $k\in[K]$, we simplify the above expression to
    \begin{align}
    \| X_k ( \Delta_0 +\Delta_k) - \varepsilon_k\|_2^2 \le \| X_k  \Delta_0 -\varepsilon_k\|_2^2 + \lambda^\prime(\|\beta_k^*\|_1-\|\hat\beta_k\|_1).
    \end{align}
    Note that this expression depends on the error of the pretraining stage $X_k \Delta_0$, for which we have established bounds. Expanding the square and simplifying the terms, we obtain
    \begin{align}
    \| X_k (\Delta_0 +\Delta_k)\|_2^2   &\le \| X_k  \Delta_0\|_2^2 -2 \Delta_0^T X_k^T\varepsilon_k + 2  (\Delta_0 +\Delta_k)^T X_k^T \varepsilon_k\nonumber\\
    &\qquad\qquad+ \lambda^\prime(\|\beta_k^*\|_1-\|\hat\beta_k\|_1)\nonumber\\
    &=\| X_k  \Delta_0\|_2^2  +  2\Delta_k^T X_k^T \varepsilon_k + \lambda^\prime(\|\beta_k^*\|_1-\|\hat\beta_k\|_1).
    \end{align}
    We sum the left-hand-side for $k\in [K]$ and obtain.
    \begin{align*}
    \sum_{k=1}^K \| X_k (\Delta_0 +\Delta_k)\|_2^2 =\sum_{k=1}^{K} \left(\| X_k  \Delta_0\|_2^2  +  2\Delta_k^T X_k^T \varepsilon_k + \lambda^\prime(\|\beta_k^*\|_1-\|\hat\beta_k\|_1)\right).
    \end{align*}
    We use the bound $2\sum_k \Delta_k^T X_k^T \varepsilon_k\le 2\sum_k \|\Delta_k\|_1 \|X_k^T \varepsilon_k\|_{\infty}\le  2 \sum_k (\|\beta^*\|_1+\|\hat\beta\|_1) \max_{k\in[K]} \| X_k^T\varepsilon_k\|_\infty$, where we applied the Cauchy Schwarz inequality twice. Next, we let $\lambda^\prime \ge 2\max_{k\in[K]} \| X_k^T\varepsilon_k\|_\infty$ in order to cancel terms involving $\|\hat \beta_k\|_1$, we obtain
    \begin{align*}
    \sum_{k=1}^K \| X_k (\Delta_0 +\Delta_k)\|_2^2 =\sum_{k=1}^{K} \| X_k  \Delta_0\|_2^2  +  2\sum_k\|\beta_k^*\|_1 \max_{k\in[K]} \| X_k^T\varepsilon_k\|_\infty + \lambda^\prime \sum_k\|\beta_k^*\|_1.
    \end{align*}
    Following the concentration bound for the maximum of sub-Gaussian variables as in the analysis of the pretraining stage, $\|X_k^T\varepsilon_k\|_{\infty}$ is bounded by $2\sigma \sqrt{n/K}\sqrt{\log(pK)}$ with high probability for all $k\in[K]$.
We combine  the above bound with the error on the pretraining stage in \eqref{eqn:error_pretrain}, and obtain
\begin{align}
\frac{1}{n} \sum_{k=1}^K \| X_k (\Delta_0 +\Delta_k)\|_2^2 &\le \|\beta_0^*\|_1\left(\frac{4 C \sigma \sqrt{\log(p)}}{\sqrt{n}} + \frac{2 C^\prime \sum_k \|\beta_k^*\|_1}{K}+\frac{\lambda}{n} \right) \nonumber \\
&\quad\quad +  \sum_k\|\beta_k^*\|_1 \left(\frac{4\sigma\sqrt{\log(pK)}}{\sqrt{nK}} + \frac{\lambda^\prime}{n}\right)\,,
\end{align}
with probability at least $1-C_3/n$ when the regularization parameters satisfy $\lambda \ge \sqrt{n}{2\|\beta_0^*\|_1(2 C \sigma \sqrt{\log(p)}} +  C^\prime \frac{n}{K}\sum_k \|\beta_k^*\|_1)$ and $\lambda^\prime \ge 4\sigma \sqrt{n/K}\sqrt{\log(pK)}$.
\qed

\subsection{Does cross-validation work here?}
\label{sec:crossval}
In lasso pretraining,  ``final'' cross-validated error that we use for the estimation of both $\lambda$ and $\alpha$ is the error
reported in the last application of {\tt cv.glmnet}. There are many reasons
why this estimate might be biased for the test-error. As with usual cross-validation with $k$ folds, each training
set has $n-n/k$ observations (rather than $n$ and hence the CV estimate will be biased upwards. On the other hand,
in the pretrained lasso, we re-used the data in the applications of {\tt cv.glmnet}, and this should cause a downward bias.
Note that we could instead do proper cross-validation--- leaving out data and running the entire pipeline for each fold. But this would be prohibitively slow.

\begin{figure}
    \centering
    \includegraphics[width=4in]{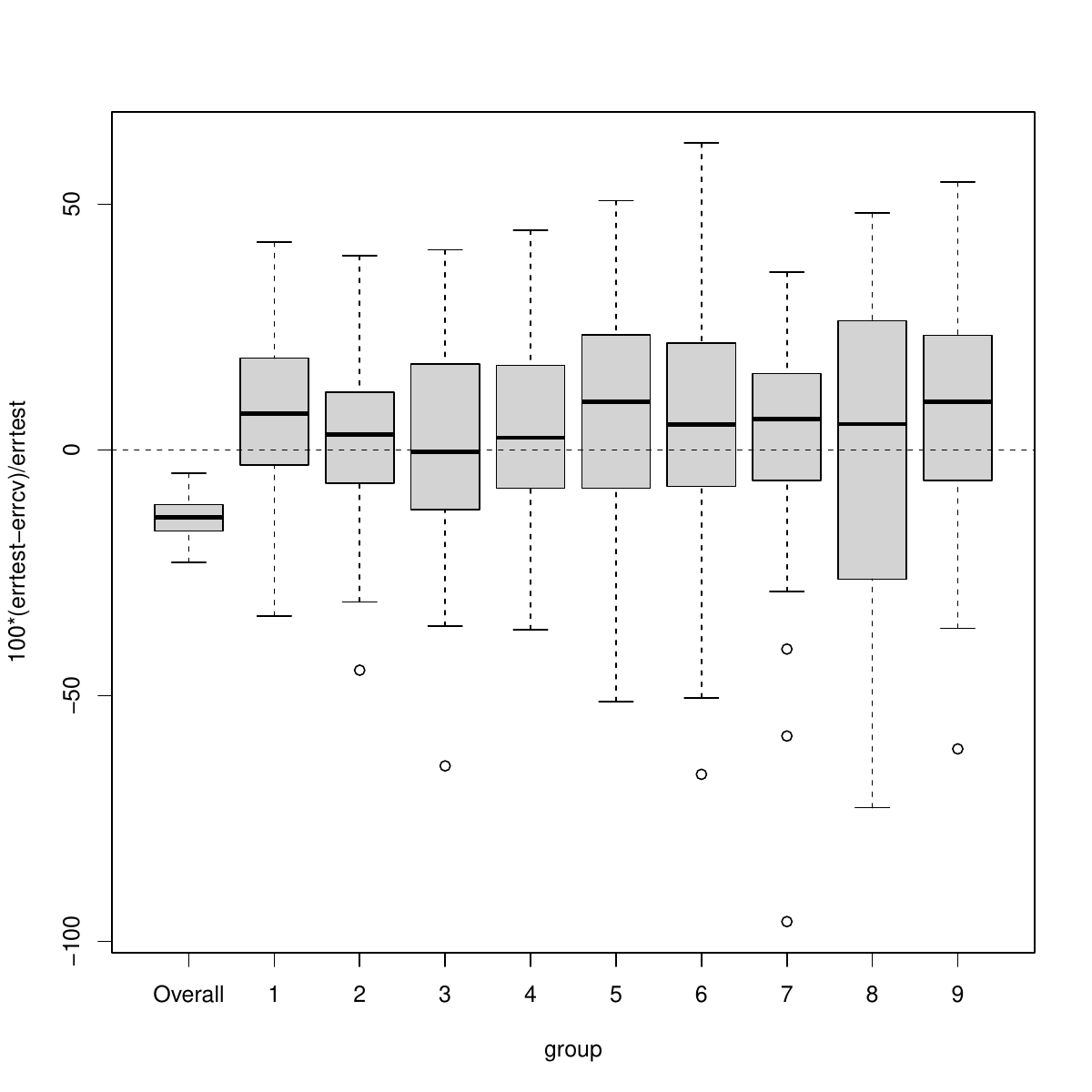}
    \caption{\em  Relative error of 5-fold CV error, as an estimate of test error (50 simulations).  Left boxplot is for the overall model; other boxplots show results for the 9 groups in pretrained lasso.
    Cross-validation overestimates the test error in the overall model, but underestimates it each of the 9 groups at the end of the two-step pretraining. However, the bias is relatively small in each case.}
    \label{fig:studyCV}
\end{figure}

We ran a simulation experiment to examine this bias. The results are in Figure \ref{fig:studyCV}.
The y-axis shows the relative error in the CV estimate as a function of the true test error.
The boxplot on the left corresponds to the overall model fit via the lasso: as expected, the estimate is a little biased upwards. The other boxplots show that the final reported CV error is on the order of 5 or 10\% too small as an estimate of the test error,
Hence this bias does not seem like a major practical problem, but should be kept in mind.

The data were simulated with ${n=500, p=1000, {\rm SNR}=2.28}$, $K=9$ groups. The class sizes are as follows: there is one group each with $n = 100$ and $n = 80$; two groups with $n = 60$, and 5 groups with $n = 40$. The shared component, $\beta_0$, is $1$ for the first 10 coefficients and $0$ otherwise. In each of the $9$ groups, the individual coefficients modify these coefficients: $\{\beta_{1j}, \beta_{2j}, \dots, \beta_{9j}\} = \{19, 16, 14, 11, 9, 7, 4, 2, 0\}$ for $j$ in 1 to 10. Additionally, the remaining values $\beta_{kj}$ for $j > 10$ are $0$ or $1$ with 9 features having the value $1$, and the nonzero entries of the $\beta_k$s are non-overlapping.  A test set of size 5000 was also generated in the same way.

\end{appendices}

\end{document}